\documentclass[12pt,a4paper]{article}

\usepackage{amsmath}
\usepackage{amssymb}
\usepackage{hyperref}
\usepackage{graphics}

\let\oldappendix=\appendix
\let\oldsection=\section
\renewcommand{\appendix}{\oldappendix%
\def\theequation{\Alph{section}.\arabic{equation}}%
\renewcommand{\section}{\setcounter{equation}{0}\oldsection}}

\newcommand{\PI}{\pi}

\newcommand{\I}{i}
\newcommand{\sfrac}[2]{{\textstyle\frac{#1}{#2}}}
\newcommand{\mmass}[2][2]{{m_{#2}^{#1}}}
\newcommand{\mmean}[1][2]{\mmass[#1]{8}}
\newcommand{\mtop}[1][2]{\mmass[#1]{0}}
\newcommand{\mom}[2][2]{p^{#1}_{#2}}
\newcommand{\bonbon}[2]{I_{#2}(#1)}
\newcommand{\tad}[1]{\Delta_{#1}}
\newcommand{\tsum}{\mathop{\textstyle\sum}}
\newcommand{\decay}{f}
\newcommand{\decaypi}{\decay_{\pi}}
\newcommand{\coeffv}[2]{v_{#1}^{(#2)}}
\newcommand{\cbeta}[2]{\beta_{#1}^{(#2)}}
\newcommand{\betaSU}[1]{\beta_{#1}^{\scriptscriptstyle SU(3)}}
\newcommand{\cvtwid}[2]{\tilde{v}_{#1}^{(#2)}}
\newcommand{\cbtwid}[2]{\tilde{\beta}_{#1}^{(#2)}}
\newcommand{\Lagr}{\mathcal{L}}
\newcommand{\keV}{\,\mathrm{keV}}
\newcommand{\MeV}{\,\mathrm{MeV}}
\newcommand{\GeV}{\,\mathrm{GeV}}

\newcommand{\artanh}{\mathop{\mathrm{artanh}}\nolimits}
\newcommand{\arcoth}{\mathop{\mathrm{arcoth}}\nolimits}

\newcommand{\indup}[1]{_{\scriptscriptstyle\mathrm{#1}}}

\hypersetup{
  pdfauthor={N. Beisert, B. Borasoy},
  pdftitle={The eta' -> eta pi pi decay in U(3) chiral perturbation theory},
  pdfsubject={12.39.Fe, 13.25.Jx},
  pdfkeywords={Chiral perturbation theory, QCD, infrared regularization, hadronic decays of mesons, eta'}
}

\begin{document}

\hfill 

\hfill 

\bigskip\bigskip

\begin{center}

{{\Large\bf  The $\mbox{\boldmath$\eta'$} \rightarrow 
\mbox{\boldmath$\eta \pi \pi$}$ decay in\\
$\mbox{\boldmath$U(3)$}$ chiral perturbation theory 
\footnote{Work supported in part by the DFG}}}

\end{center}

\vspace{.4in}

\begin{center}
{\large N. Beisert\footnote{email: nbeisert@physik.tu-muenchen.de}
 and  B. Borasoy\footnote{email: borasoy@physik.tu-muenchen.de}}

\bigskip

\bigskip

\href{http://www.ph.tum.de/}{Physik Department}\\
\href{http://www.tum.de/}{Technische Universit{\"a}t M{\"u}nchen}\\
D-85747 Garching, Germany \\

\vspace{.2in}

\end{center}

\vspace{.7in}

\thispagestyle{empty} 

\begin{abstract}
The dominant decay mode of the $\eta'$, $\eta' \rightarrow \eta \pi \pi$,
is investigated within the framework of infrared regularized $U(3)$
chiral perturbation theory up to fourth chiral order and including one-loop
corrections. We find reasonable agreement with experimental data and observe
convergence of the chiral series.
\end{abstract}\bigskip

\begin{center}
\begin{tabular}{ll}
\textbf{PACS:}&12.39.Fe, 13.25.Jx\\[6pt]
\textbf{Keywords:}& Chiral perturbation theory, infrared regularization, \\&
hadronic decays of mesons, $\eta'$.
\end{tabular}
\end{center}

\vfill

\section{Introduction}\label{sec:intro}

In the limit of vanishing light quark masses $m_u, m_d$ and $m_s$ the QCD 
Lagrangian exhibits a symmetry under chiral $SU(3)_L \times SU(3)_R$
transformations which is broken down spontaneously to $SU(3)_V$ giving rise
to the eight pseudoscalar Goldstone bosons $(\pi, K, \eta)$. The interactions
at low energies amongst the Goldstone bosons can be treated in a perturbative
fashion within the framework of chiral perturbation theory (ChPT).
The axial $U(1)$ anomaly of the QCD Lagrangian, on the other hand, prevents
the corresponding pseudoscalar singlet from being a Goldstone boson. Indeed,
the lightest candidate would be the $\eta'$ with a mass of $958\MeV$ which
is considerably heavier than the octet states. Therefore, in conventional
ChPT the $\eta'$ is not included explicitly, although its effects are hidden
in a contribution to a coupling coefficient of the higher order Lagrangian,
a so-called low-energy constant (LEC).

In order to describe the physical processes involving the $\eta'$, such as the 
hadronic decay modes $\eta' \rightarrow \eta \pi \pi, \pi \pi \pi$, the
anomalous decays $\eta' \rightarrow \gamma \gamma, \pi^+ \pi^- \gamma$,
as well as photoproduction of the $\eta'$ on the nucleons and $\eta'$
production in $pp$ collisions, it is necessary to include the $\eta'$ 
explicitly within the effective theory. The inclusion of the $\eta'$ can be
achieved either by introducing the topological charge density as a background 
field in the effective Lagrangian, see e.g. \cite{DiVecchia:1980ve, Witten:1980sp}, 
or by extending the chiral $SU(3)_L \times SU(3)_R$ symmetry to $U(3)_L \times U(3)_R$
\cite{Gasser:1985gg, Leutwyler:1996sa, Herrera-Siklody:1997pm}. 
(The equivalence of both schemes has been proved in 
\cite{Borasoy:2000xj}.)

Due to the large $\eta'$ mass, however, the convergence of the chiral expansion 
is questionable. As a matter of fact, the appearance of a mass
that does not vanish in the chiral limit, such as the mass of the $\eta'$
will spoil the chiral counting scheme so that higher loops will still contribute to
lower orders. This can be prevented by imposing large $N_c$ counting rules
within the effective theory in addition to the chiral counting scheme.
In the large $N_c$ limit the axial anomaly vanishes and the $\eta'$ converts 
into a Goldstone boson. The properties of the theory may then be analyzed in a 
triple expansion in powers of small momenta, quark masses and 1/$N_c$, see
e.g. \cite{Leutwyler:1996sa, Herrera-Siklody:1997pm, Kaiser:1998ds,Kaiser:2000gs}. In particular, $m_{\eta'}$ is treated as a small 
quantity. Phenomenologically, this is not the case and we will therefore treat
the $\eta'$ as a massive state leading to a situation analogous to
ChPT with baryons. Recently, a new regularization scheme -- the so-called
infrared regularization -- has been proposed which maintains Lorentz and chiral
invariance explicitly at all stages of the calculation while providing
a systematic counting scheme for the evaluation of the chiral loops \cite{Becher:1999he}.
Infrared regularization can be applied in the presence of any massive state
and has been employed in $U(3)$ ChPT in \cite{Borasoy:2001ik, Beisert:2001qb}.

In the present work, we apply this approach to the dominant decay of the 
$\eta'$: $\eta' \rightarrow \eta \pi \pi$, including one-loop corrections.
This decay has been the subject of many investigations using chiral effective 
Lagrangians, see e.g. \cite{DiVecchia:1980ve, Belkov:1987ms, Fajfer:1989ij, Akhoury:1989as, Herrera-Siklody:1999ss}, however one-loop corrections 
have not been included so far. Of particular interest is the recent work 
\cite{Herrera-Siklody:1999ss} where a systematic study up to next-to-leading order within the 
framework of large $N_c$ ChPT has been undertaken. One-loop corrections have not 
been included in \cite{Herrera-Siklody:1999ss} since they start at third order in the combined 
chiral and 1/$N_c$ expansion, and although the results at tree level were 
promising, no statement about the convergence of the expansion could be made.
If, on the other hand, large $N_c$ counting rules are not employed, loop 
corrections start already at next-to-leading order.
Our investigation will shed some light on the convergence issue of infrared
regularized $U(3)$ ChPT. It includes for the first time unitarity corrections 
whereas in \cite{Borasoy:2001ik, Beisert:2001qb} tadpoles were the only loop corrections.
The present calculation will be a crucial test for the convergence of 
the chiral series within this approach.

It has been claimed that this decay can be described in a tree-level model
via the exchange of the
scalar mesons $\sigma(560), f_0(980)$ and $a_0(980)$ which are combined
together with $\kappa(900)$ into a nonet \cite{Fariborz:1999gr} (see also \cite{Schechter:1971tc,Singh:1975aq,Deshpande:1978iv,Bramon:1980ni,Castoldi:1988dm}).
The authors find that the exchange of the scalar resonance $a_0(980)$ 
dominates. It will be interesting to see if we are able to reproduce the 
experimental data reasonably well without the explicit inclusion of resonances
such as the $a_0(980)$ although their effects might be hidden in some LECs.

In the next section we present the formalism for the 
$\eta' \rightarrow \eta \pi \pi$ decay as well as the pertinent Lagrangian
up to fourth chiral order and contributions from tree diagrams. 
The appearing chiral loop corrections are discussed
in Sec. \ref{sec:loops} and Sec. \ref{sec:compare} contains the comparison with experiment.
We summarize our findings and conclusions in Sec. \ref{sec:conclude}.

\section{Formalism and leading contributions}
\label{sec:tree}

The process $\eta' \rightarrow \eta \pi \pi$ is the dominant decay mode of the
$\eta'$. The measured rates are \cite{Groom:2000in}:
\begin{eqnarray} 
\Gamma(\eta' \rightarrow \eta \pi^+ \pi^-)  &=& (90 \pm 8)\keV 
\nonumber \\
\Gamma(\eta' \rightarrow \eta \pi^0 \pi^0)  &=& (42 \pm 4)\keV
\end{eqnarray}
with the ratio given by
\begin{equation} 
r = \frac{\Gamma(\eta' \rightarrow \eta \pi^+ \pi^-)}
         {\Gamma(\eta' \rightarrow \eta \pi^0 \pi^0)} =  2.1 \pm 0.4
\end{equation}
which is consistent with the ratio of 2 in the isospin limit.
Isospin violating corrections to the decay rates seem to be small and will
therefore be neglected throughout this work. Taking this into account the 
amplitudes  of the two decay channels are equal
\begin{equation} 
A_{\eta' \rightarrow \eta \pi^+ \pi^-}(s,t,u) =
A_{\eta' \rightarrow \eta \pi^0 \pi^0}(s,t,u) 
\end{equation}
where we introduced the Mandelstam variables
\begin{equation} 
s = (p_{\eta'} - p_{\eta})^2, \qquad
t = (p_{\eta'} - p_{\pi1})^2, \qquad
u = (p_{\eta'} - p_{\pi2})^2 
\end{equation}
with $p_{\varphi}$ being the four-momentum of the meson $\varphi$.
In order to calculate the amplitude, we restrict ourselves to the effective
Lagrangian up to next-to-leading order, i.e. to fourth order in the chiral 
expansion. The construction of the pertinent Lagrangian has been given 
elsewhere, see e.g. \cite{Herrera-Siklody:1997pm, Kaiser:1998ds,Kaiser:2000gs, Borasoy:2001ik}, and we sketch the results briefly.
Note that 1/$N_c$ counting rules are not necessary to set up the Lagrangian
including the $\eta'$.
The effective Lagrangian for the pseudoscalar meson nonet
($\pi, K, \eta_8, \eta_0$) reads up to second order in the derivative expansion
\begin{equation}  \label{eq:mes1}
\Lagr^{(0+2)} = - v_0(\eta_0) + \tsum\nolimits_i v_i(\eta_0)\, O_i^{(2)}
\end{equation}
with the relevant second order operators
\begin{equation}
O_1^{(2)}=-\langle C^\mu C_\mu \rangle,
\quad
O_2^{(2)}=\langle M \rangle,
\quad
O_3^{(2)}=\langle i N \rangle,
\quad
O_4^{(2)}=\langle C^\mu\rangle\langle C_\mu \rangle.
\end{equation}
We used the abbreviations $C_\mu=U^\dagger \partial_\mu U$, 
$M=U^\dagger\chi+\chi^\dagger U$, $N=U^\dagger\chi-\chi^\dagger U$, and
$U$ is a unitary $3 \times 3$ matrix containing the meson nonet.
Its dependence on $\eta_8$ and $\eta_0$ is given by 
\begin{equation}\label{eq:norm}
U=\exp\bigl(\mathrm{diag}(1,1,-2)\cdot i\eta_8/ \sqrt{3}\decay
+i\sqrt{2}\eta_0/\sqrt{3}\decay+\ldots\bigr).
\end{equation}
The expression $\langle \ldots \rangle$ denotes 
the trace in flavor space, $\decay$ is the pseudoscalar 
decay constant in the chiral 
limit and the quark mass matrix $\mathcal{M} = \mbox{diag}(m_u,m_d,m_s)$
enters in the combination  $\chi  = 2 B \mathcal{M} $
with $B=-\langle 0|\bar{q}q|0\rangle/\decay^2$ being the order
parameter of the spontaneous symmetry violation.
In the present investigation we work in the isospin limit
$m_u=m_d=\hat{m}$ and, therefore, only $\eta$-$\eta'$ mixing occurs.
We have neglected terms which do not contribute to the process considered here, 
such as those containing external sources.

The coefficients $v_i$ are functions of the singlet field 
and are expanded in terms of $\eta_0$
\begin{equation}\label{eq:vexpand}
v_i(\eta_0) =
  \coeffv{i}{0} 
+ \coeffv{i}{1} \frac{\eta_0}{\decay} 
+ \coeffv{i}{2} \frac{\eta_0^2}{\decay^2} 
+ \coeffv{i}{3} \frac{\eta_0^3}{\decay^3}
+ \ldots
\end{equation}
At a given order of derivatives of the meson fields $U$ and insertions 
of the quark mass matrix  $\mathcal{M}$ one obtains an infinite string of 
increasing powers of $\eta_0$ with couplings $\coeffv{i}{j}$ which are not fixed by chiral symmetry
and have to be determined by phenomenologically.
Parity conservation implies that the $v_i$ are all even functions
of $\eta_0$ except $v_3$, which is odd.
The correct normalization for the quadratic terms of the mesons is obtained by 
$v_1(0) = v_2(0) = v_1(0)-3v_4(0)=\frac{1}{4}\decay^2$.

The next-to-leading order Lagrangian reads
\begin{equation}
\Lagr^{(4)}=\tsum\nolimits_i \beta_i(\eta_0)\, O_i
\end{equation}
with contributing fourth order operators 
\begin{equation}
\begin{array}{ll}
O_{\phantom{0}0}=\langle C^\mu C^\nu C_\mu C_\nu\rangle,&
O_{\phantom{0}1}=\langle C^\mu C_\mu\rangle\langle C^\nu C_\nu\rangle,\\
O_{\phantom{0}2}=\langle C^\mu C^\nu\rangle\langle C_\mu C_\nu\rangle,&
O_{\phantom{0}3}=\langle C^\mu C_\mu C^\nu C_\nu\rangle,\\
O_{13}=-\langle C^\mu\rangle\langle C_\mu C^\nu C_\nu\rangle,&
O_{14}=-\langle C^\mu\rangle\langle C_\mu\rangle \langle C^\nu C_\nu\rangle,\\
O_{15}=-\langle C^\mu\rangle\langle C^\nu\rangle \langle C_\mu C_\nu\rangle,\qquad&
O_{16}=\langle C^\mu\rangle\langle C_\mu\rangle\langle C^\nu\rangle\langle C_\nu\rangle,\\
O_{\phantom{0}4}=-\langle C^\mu C_\mu\rangle\langle M\rangle,&
O_{\phantom{0}5}=-\langle C^\mu C_\mu M\rangle,\\
O_{17}=\langle C^\mu \rangle\langle C_\mu\rangle\langle M\rangle,&
O_{18}=-\langle C^\mu \rangle\langle C_\mu M\rangle,\\
O_{21}=\langle C^\mu C_\mu i N\rangle,&
O_{22}=\langle C^\mu C_\mu\rangle\langle i N\rangle,\\
O_{23}=\langle C^\mu \rangle\langle C_\mu i N\rangle,&
O_{24}=\langle C^\mu \rangle\langle C_\mu\rangle\langle i N\rangle,\\
O_{\phantom{0}6}=\langle M\rangle\langle M\rangle,&
O_{\phantom{0}7}=\langle N\rangle\langle N\rangle,\\
O_{\phantom{0}8}=\sfrac{1}{2}\langle MM+NN\rangle,&
O_{12}=\sfrac{1}{4}\langle MM-NN\rangle,\\
O_{25}=\langle iMN\rangle,&
O_{26}=\langle M\rangle\langle iN\rangle,
\end{array}
\end{equation}
where we kept the notation from \cite{Herrera-Siklody:1997pm}.
The coefficients $\beta_i$ are expanded in $\eta_0$ in the same manner 
as the $v_{i}$ in \eqref{eq:vexpand}; they are even (odd) functions of $\eta_0$
for an even (odd) number of operators $N$. 
It turns out to be more convenient to include the $\beta_0$ term, although
there is a Cayley-Hamilton matrix identity that enables one to 
remove this term leading to modified coefficients 
$\beta_i$, $i=1,2,3,13,14,15,16$ \cite{Herrera-Siklody:1997pm}.
Here we do not make use of the Cayley-Hamilton identity
and keep all couplings in order to present the most general expressions
in terms of these parameters. One can then drop one of the 
$\beta_i$ involved in the Cayley-Hamilton identity
at any stage of the calculation.

Having presented the effective Lagrangian up to fourth chiral order we now
proceed by calculating tree diagrams contributing to this decay.
The contribution from $\Lagr^{(2)}$ reads
\begin{equation}   \label{eq:a2tree}
A_2=\frac{4\sqrt{2}\,\cvtwid{2}{1} \mmass{\pi}}{3 \decaypi^4}
\end{equation}
with $m_{\pi} = 138$ MeV being the average pion mass,
$\decaypi= 92.4$ MeV the pion decay constant and
the abbreviation
\begin{equation}   \label{eq:cvt31}
\cvtwid{2}{1} = \sfrac{1}{4}\decay_\pi^2-\sfrac{1}{2}\sqrt{6}\coeffv{3}{1}.
\end{equation}
Inserting the value $\cvtwid{2}{1} = 2.67 \times 10^{-3}\GeV^2$, 
obtained from a fit to the pseudoscalar decay constants \cite{Beisert:2001qb},
leads to a decay width of $\Gamma=1.9 \keV$ 
which is about
20 times smaller than the experimental value. 
This has already been pointed out 
in \cite{Fajfer:1989ij, Akhoury:1989as, Herrera-Siklody:1999ss},
where an even smaller result is obtained due to the omission of $\coeffv{3}{1}$.
Note that we have replaced the pseudoscalar decay constant $\decay$ 
appearing in the Lagrangian by the physical parameter $\decaypi$ which 
coincides with $\decay$ in the chiral limit. The difference in the decay 
amplitude is of higher order and is included in the higher chiral order
contributions, see below. We proceed in a similar way by substituting the 
physical pion and kaon masses for the corresponding masses at leading order
and writing the differences into the higher order contributions. 
An important feature of \eqref{eq:a2tree} is that 
$\eta$-$\eta'$ mixing does not show up at this order
and, as a matter of fact, the amplitude calculated 
in \eqref{eq:a2tree} corresponds 
to the decay $\eta_0\to\eta_8\pi\pi$. Within the used scheme 
$\eta$-$\eta'$ mixing is of second chiral order \cite{Beisert:2001qb} and
the corrections due to mixing on the decay amplitude \eqref{eq:a2tree} will
be of fourth chiral order. 

Apparently, tree level diagrams from the Lagrangian $\Lagr^{(2)}$ are not 
sufficient to obtain reasonable agreement with experiment. One must therefore
go beyond the leading order Lagrangian and
take into account tree diagrams from the Lagrangian $\Lagr^{(4)}$ and loops
with vertices from $\Lagr^{(2)}$. The loop contributions will be discussed in
detail in the next section and we will start here by evaluating tree diagrams
from $\Lagr^{(4)}$. The complete tree level amplitude $A_4$ 
from $\Lagr^{(4)}$ is given in App. \ref{sec:ampli}.
Comparing the tree level contributions one makes the following crucial 
observation: although, strictly speaking, of fourth chiral order in the
expansion in the derivatives and meson masses, the terms with the LEC 
combinations $\beta_{0,3,13}, \beta_{5,18},\cbtwid{4}{1}$ and $\cbtwid{5}{1}$ 
(see App. \ref{sec:b} for the definitions) contribute already at lower 
orders if one makes use of $\mom{\eta'}=\mmass{\eta'} \sim \mathcal{O}(1)$. 
More generally, we will consider the Mandelstam variables $s, t$ and $u$ as
zeroth chiral order. 
This counting is motivated by the fact that in the 
chiral limit $s, t$ and $u$ range from 0 to $m_{\eta'}^2$.
From that point of view, it is more convenient to summarize the terms
$\cvtwid{2}{1}, \beta_{0,3,13}, \beta_{5,18}, \cbtwid{4}{1}$ and $\cbtwid{5}{1}$ in
\begin{eqnarray}\label{eq:a02}
A\indup{LO} =&&  \frac{4\sqrt{2}}{3 \decaypi^4}  \cvtwid{2}{1} \mmass{\pi}
- \frac{8\sqrt{2}}{3 \decaypi^4}  \beta_{5,18} \mmass{\pi} \mmass{\eta'}
\nonumber \\ && \mathord{}
+ \frac{4\sqrt{2}}{3 \decaypi^4}  \beta_{0,3,13}  
(s^2 + t^2 + u^2 -2\mmass[4]{\pi}-\mmass[4]{\eta}-\mmass[4]{\eta'})
\nonumber \\ && \mathord{}
+\frac{16\sqrt{2}\,(\mmass{K}-\mmass{\pi})\cbtwid{4}{1}}{3\decaypi^4}
\big(2\mmass{\pi}-s\big)
+\frac{4\sqrt{2}\,\mmass{\pi}\cbtwid{5}{1}}{3\decaypi^4}
(2\mmass{\eta'}-s-t-u)
\nonumber \\ && \mathord{}
+R_{8\eta'}^{(2)} A_{88\pi\pi,\beta_{0}\ldots\beta_{3}}
+R_{0\eta}^{(2)}
\frac{4\coeffv{1}{2}}{\decaypi^4}\big(2\mmass{\pi}-s\big)
\end{eqnarray}
and to distinguish them from the remaining terms 
in $A_{4}$. The other contributions from the fourth order Lagrangian
which have not been included in $A\indup{LO}$ \eqref{eq:a02} are all proportional
to $\mmass[4]{\varphi}$ with $\varphi=\pi,K,\eta$ and therefore 
contribute to the decay amplitude at fourth chiral order. 
This clearly separates the terms given in \eqref{eq:a02} from the 
remaining tree level contributions in App. \ref{sec:ampli}. 
The last two terms are due to $\eta$-$\eta'$ mixing: $A_{88\pi\pi,\beta_0\ldots\beta_3}$ denotes 
the contributions of the amplitude $A_{88\pi\pi}$
in App. \ref{sec:ampli} proportional
to $\cbeta{0}{0}+\cbeta{3}{0}$, $\cbeta{1}{0} $ and $\cbeta{2}{0}$, and
$R_{8\eta'}^{(2)}, R_{0\eta}^{(2)}$ describe the mixing of the $\eta$ and $\eta'$ fields 
at one-loop order \cite{Beisert:2001qb} with
\begin{eqnarray}
R^{(2)}_{8\eta'}&=&\sfrac{8}{3}\sqrt{2}\,(\mmass{K}-\mmass{\pi})
\bigl(2\mtop\beta_{5,18}-\cvtwid{2}{1}\bigr)\big/ \big(\decaypi^2\mtop\big),
\nonumber\\
R^{(2)}_{0\eta}&=&\sfrac{8}{3}\sqrt{2}\,(\mmass{K}-\mmass{\pi}) \cvtwid{2}{1}\big/\big(\decaypi^2\mtop\big),
\nonumber\\
R^{(2)}_{8\eta}&=&\bigl(-12\mmean\cbeta{4}{0}-4\mmass{\eta}\cbeta{5}{0}
+\sfrac{1}{2}\tad{K}\bigr)/\decaypi^2,
\nonumber\\
R^{(2)}_{0\eta'}&=&-4\mmean\beta_{4,5,17,18}/\decaypi^2,
\end{eqnarray}
where 
$\mtop=2\coeffv{0}{2}/\decaypi^2$ 
and $\mmean=\sfrac{2}{3}\mmass{K}+\sfrac{1}{3}\mmass{\pi}$.
The chiral logarithm $\tad{K}$ is given in \eqref{eq:chilog}.
We observe that the contribution proportional to
$\beta_{0,3,13}$ is the only term in $A\indup{LO}$ that does not vanish for
zero quark masses; 
for reasons to be explained below it is expected to dominate.

In order to obtain a rough estimate for the tree diagram contributions, we will
borrow the values for the coefficients $\cbeta{i}{0}$ from conventional $SU(3)$ ChPT
\cite{Bijnens:1994qh}. 
There is, however, one complication: in $SU(3)$ ChPT
the operator $O_0$ is usually eliminated by making use of a Cayley-Hamilon 
identity which changes the 1/$N_c$ hierarchy of some of the 
LECs $L_i$ \cite{Herrera-Siklody:1997pm}.
We prefer to keep it explicitly but
must transform the values of the $L_i$ accordingly. 
An estimate of the coupling of $O_0$ that respects the $1/N_c$ hierarchy is 
given by QCD bosonization models 
\cite{Balog:1984ps,Andrianov:1985ay,Espriu:1990ff,Bijnens:1993uz}
\begin{equation}
\betaSU{0} = \frac{N_c}{192 \pi^2} \approx 1.58 \times 10^{-3},
\end{equation}
In order to introduce the $\betaSU{0}$ term
within the $SU(3)$ framework one must use the transformed 
couplings $\betaSU{i}$ as follows
\begin{equation}
\betaSU{1}=L_1-\sfrac{1}{2} \betaSU{0} , \quad
\betaSU{2}=L_2-\betaSU{0} , \quad
\betaSU{3}=L_3+2\betaSU{0} , \quad
\betaSU{i}=L_i \mbox{ for }i\geq 4.
\end{equation}
Note that $\betaSU{1}$ and $\betaSU{2}$ are both considerably 
closer to zero than $L_1$ and $L_2$ in agreement with large $N_c$ rules.
The values $\cbeta{i}{0}$ can be taken directly from the transformed values of
$SU(3)$ ChPT with the exception of $\cbeta{7}{0}$.
The value of $\betaSU{7}$ is entirely saturated by the exchange of an $\eta'$
\cite{Ecker:1989te}. Starting with $\cbeta{7}{0} =0 $ in the extended
framework and integrating out the $\eta'$ field produces the value of 
$\betaSU{7}$ \cite{Beisert:2001qb}. Hence we set
\begin{equation}
\cbeta{i}{0}=\betaSU{i} \mbox{ for }i\neq 7, \quad
\cbeta{7}{0}=0.
\end{equation}
Having fixed the LECs which already appear in conventional 
$SU(3)$ ChPT we can now proceed in estimating the remaining
$\cbeta{i}{0}$ with $i\geq 13$. To this end, we notice that these new unknown
couplings enter only in combinations with the known $\cbeta{i}{0}$, 
$i=0,\ldots,8$, e.g. $\cbeta{13}{0}$ and $\cbeta{18}{0}$ appear in
$\beta_{0,3,13}$ and $\beta_{5,18}$, respectively,
where they represent OZI violating corrections
(for $\beta_{13}$ this is only true if 
the Cayley-Hamilton identity is used as described above).
Therefore, we make the rough estimate by neglecting the unknown OZI 
suppressed couplings within these parameter combinations and keeping
only the known parameters.
For the two cases mentioned this yields 
$\beta_{0,3,13}\approx \cbeta{0}{0}+\cbeta{3}{0}$
and $\beta_{5,18}\approx \cbeta{5}{0}$.
Furthermore,  $\cbeta{17}{0}$ and $\cbeta{18}{0}$ are OZI violating 
corrections to $\cbeta{5}{0}$ and we use the estimate 
$\beta_{4,5,17,18}\approx 3 \cbeta{4}{0} + \cbeta{5}{0}$
This rough estimate for the parameter combinations is motivated by conventional 
$SU(3)$ ChPT, where a decent estimate of the LECs can be obtained via
resonance exchange if
resonance couplings are used which obey the OZI rule \cite{Ecker:1989te}. 
Contributions from OZI suppressed couplings to the operators in $SU(3)$ ChPT
seem to be negligible and we will assume the same also in $U(3)$ ChPT. 

Finally, the parameters $\beta_{21},\beta_{22}$ and $\beta_{25},\beta_{26}$
from parity-odd operators enter the calculation. The coupling $\beta_{21}$
appears together with $\beta_{5}$, while $\beta_{22}$ appears with $\beta_{4}$.
Both $\beta_{21}$ and $\beta_{22}$ are suppressed by one power of 1/$N_c$ in
comparison with $\beta_{4}$ and $\beta_{5}$, respectively. The same applies for
$\beta_{25}$ ($\beta_{26}$) which combines with $\beta_{8}$ 
($\beta_{6}+\beta_{7}$). We will also drop these 1/$N_c$ corrections from parity-odd 
operators, hence generalizing our approximation for the OZI violating 
contributions.
In particular, $\cbtwid{4}{1}$ and $\cbtwid{5}{1}$ are approximated by 
$\beta_{4}$ and $\beta_{5}$, respectively.
Making use of the 1/$N_c$ hierarchy for the LECs, we are only left with known
parameters $\cbeta{i}{0}$ of the fourth order Lagrangian. 

The same estimate of dropping $1/N_c$ corrections in parameter combinations
can also be applied to coefficients from the lowest order Lagrangian.
Keeping $\cvtwid{2}{1}=2.67\times 10^{-3}\GeV^2$ fixed at its
value determined from a fit to 
the pseudoscalar decay constants \cite{Beisert:2001qb}
$\coeffv{2}{2}$ enters only in the combination
$\cvtwid{2}{2}$ where it is $1/N_c$ suppressed. It will thus be neglected,
yielding the estimate 
$\cvtwid{2}{2}\approx\coeffv{2}{0}-\sqrt{6}\coeffv{3}{1}\approx
3.2\times 10^{-3}\GeV^2$.
We also omit the coefficient $\coeffv{1}{2}$ which should have
a small value since it is
$1/N_c$ suppressed with respect to $\coeffv{1}{0}= f^2/4$.
Indeed, small variations of $\coeffv{1}{2}$ within a realistic range
have shown that this contribution has almost no effect on the results.
Dropping $1/N_c$ suppressed corrections
simplifies the parameter combinations considerably and reduces them to the
contribution of the leading $N_c$ piece. Afterwards we are mainly left with 
known parameters from conventional $SU(3)$ ChPT.
Of course, one is free to keep the $1/N_c$ corrections, but our results
will lose most of their predictive power and it will be relatively easy to
accommodate the experimental data by finetuning these LECs. 
The neglection of such couplings provides a more stringent
test of our approach.
Estimating the LEC combinations in the
amplitude by the contributions from the leading $N_c$ pieces
should not be confused with imposing large $N_c$ counting rules
in ChPT. We do not make use of large $N_c$ rules in the
calculation of the amplitude which is obtained by applying infrared 
regularized $U(3)$ ChPT. 

Values for the LECs in $SU(3)$ ChPT were given in \cite{Bijnens:1994qh}.
They correspond to the values 
$\cbeta{0}{0}=1.6\pm 0.0$,
$\cbeta{1}{0}=-0.4\pm 0.3$,
$\cbeta{2}{0}=-0.2\pm 0.3$,
$\cbeta{3}{0}=-0.3\pm 1.1$,
$\cbeta{4}{0}=-0.3\pm 0.5$,
$\cbeta{5}{0}=1.4\pm 0.5$,
$\cbeta{6}{0}=-0.2\pm 0.3$,
$\cbeta{7}{0}=0.0\pm 0.2$ and
$\cbeta{8}{0}=0.9\pm 0.3$
(in units of $10^{-3}$)
in $U(3)$ ChPT when applying the procedure described above. 
All the other LECs are approximately zero and will be neglected.
The three combinations of LECs in App. \ref{sec:b} have the values
$\beta_{0,3,13}=1.2\pm 1.1$, 
$\beta_{5,18}=1.4\pm 0.5$ and
$\beta_{4,5,17,18}=0.5\pm 1.6$.

We will give amplitudes at a central point in phase space,
the Mandelstam variables being given by their phase space expectation
values 
\begin{equation}\label{eq:center}
s_0=0.12\GeV^2,\quad t_0=u_0=0.567\GeV^2.
\end{equation}
The phase space of the decay is rather small, 
so that fluctuations of the amplitude have only 
little impact on the decay width and
incidentally the squared amplitude at this point is a good approximation for the decay 
width in units of $\keV$.


In order to obtain numerical results, we first note that the 
amplitude is dominated by the $\beta_{0,3,13}$ term. 
The huge uncertainty in the value of $\beta_{0,3,13}$, which is as
large as the value itself, makes a numerical prediction of the
decay width impossible, it only allows for an upper bound of roughly $200\keV$.
Hence we will use the experimental decay width and the slope 
parameter $\alpha$ (see below) to fit some of the LECs
and reduce the uncertainty of $\beta_{0,3,13}$. We find that the values
$\cbeta{0}{0}+\cbeta{3}{0}=1.1$, $\cbeta{4}{0}=-0.7$ and $\cbeta{5}{0}=1.9$ 
(in units of $10^{-3}$), which lie inside their uncertainties given above, 
reproduce the data well.
Alternatively, one could employ the central values but finetune the OZI
violating corrections which we have neglected initially.
The contributions from 1/$N_c$ suppressed parameters are then within the given
error ranges for the known LECs and our initial assumption of omitting
the 1/$N_c$ corrections seems to be justified.
Our estimate for the $A\indup{LO}$ amplitude reads then
\begin{equation} \label{eq:estimate}
A\indup{LO}(s_0,t_0,u_0) = -6.4.
\end{equation}
For comparison the decay width integrated over phase space is 
\begin{equation}
\sfrac{1}{2}\Gamma(\eta' \rightarrow \eta \pi^+ \pi^-) =
\Gamma(\eta' \rightarrow \eta \pi^0 \pi^0)
=42.2\keV.
\end{equation}
which coincides numerically with 
$|A\indup{LO}(s_0,t_0,u_0)|^2\approx (6.4)^2 \approx 40.2$.
Breaking $A\indup{LO}$ down into its constituents one obtains
\begin{equation} \label{eq:constit}
A\indup{LO}(s_0,t_0,u_0)=1.3-1.7-7.8+1.3+0.5+0.0
\end{equation}
with the $\cvtwid{2}{1}, \beta_{5,18}, 
\beta_{0,3,13}, \cbtwid{4}{1},\cbtwid{5}{1}$ and $A^{(0)}_{4,88\pi\pi}$ 
terms in order.
Apparently, the $\beta_{0,3,13}$ term dominates the amplitude being an order
of magnitude larger than $\cvtwid{2}{1}$ and $ \beta_{5,18}$. In previous work
it was realized that the only contribution to the decay from the lowest order 
Lagrangian $\mathcal{L}^{(2)}$ is given by $\cvtwid{2}{1}$ yielding a width
much smaller than the measured one \cite{Belkov:1987ms, Fajfer:1989ij, Akhoury:1989as, Herrera-Siklody:1999ss}. This dilemma
was usually cured by adding momentum dependent terms of the type 
$\beta_{0}\ldots\beta_{3}$ \cite{Belkov:1987ms, Akhoury:1989as}; within the framework of large $N_c$
ChPT the $\beta_{0}$ term turned out to be dominating although it is a
next-to-leading order effect \cite{Herrera-Siklody:1999ss}.

After fitting the value of $\beta_{0,3,13}$ to the decay width
we may reduce its uncertainty to
\begin{equation}\label{eq:b0est}
\beta_{0,3,13}=(1.1\pm 0.4)\times 10^{-3}.
\end{equation}
The error accounts for the uncertainties of both the decay widths
and the other LECs. 

The dominance of momentum dependent terms can be easily understood by the
requirement of Adler zeros.
Assuming analyticity  in the Mandelstam variables $s, t$ and $u$ (which is
certainly the case for tree diagrams), the amplitude can be expanded as
\begin{equation} \label{eq:series}
A(t,u) = a + b (t+u) + c (t+u)^2 + d (t-u)^2 + \mbox{higher orders in } t,u
\end{equation}
where we made use of the symmetry under the exchange of the two pions.
Since for $p_\eta \rightarrow 0$, i.e. $ s= m_{\eta'}^2, t = m_\pi^2 , 
u=  m_\pi^2 $, the amplitude vanishes, $a$ must be proportional to 
$m_\pi^2$. For physical values of the Mandelstam variables, on the other hand,
one has the constraint  $t+u > m_{\eta'}^2$. From the ratio 
$(m_\pi/m_{\eta'})^2 \approx 0.02$ it becomes obvious that in the physically 
allowed region momentum dependent terms might be important and even dominate
the leading order contribution $a$.

One could now argue that higher order terms with an increasing number of 
derivatives on the singlet field $\eta_0$ will lead to contributions of the 
type $(t+u)^n$ and that consequently the series in \eqref{eq:series} will not 
converge. To this end, we would like to take a closer look at terms with more 
derivatives on the singlet field $\eta_0$. Neglecting small corrections from 
$\eta$-$\eta'$ mixing, possible tree level contributions have the generic form
$\partial^k \eta_0 \partial^l \eta_8 \partial^m \pi \partial^n \pi$   
where we omitted Lorentz indices for brevity. Derivatives acting on one field 
cannot have the same Lorentz index since such terms are eliminated by using
the equation of motion for the mesons. This means that derivatives on $\eta_0$
must be contracted with derivatives on a Goldstone boson which are suppressed
by one order in the Goldstone boson masses.
Hence, we expect the series to converge, albeit slowly. In order to clarify this issue the 
inclusion of terms from the sixth order chiral Lagrangian, in particular those
with six derivatives, would be very helpful. However, at that order two-loop
diagrams contribute together with a proliferation of many new counterterms. 
Such an analysis is far beyond the scope of the present work. We will 
therefore restrict ourselves to the Lagrangian up to fourth chiral order,
assuming that corrections from higher orders in the derivative expansion are
well-behaved.

\section{Chiral loops}  \label{sec:loops}
To our knowledge, the calculation of chiral loops has been omitted in previous
work. This is due to the fact that the $\eta'$ propagation within a loop 
introduces a hadronic mass scale, $m_{\eta'}$, and spoils the chiral
counting scheme if dimensional regularization is employed for the evaluation
of the integrals. This can be prevented using large $N_c$ ChPT and a first step
towards this direction has been undertaken by calculating the next-to-leading 
order  tree diagram corrections \cite{Herrera-Siklody:1999ss}. Within this scenario loop 
diagrams start contributing at third order in the combined chiral and 1/$N_c$
expansion. Although the tree level results were quite promising, it remains
to be seen whether loop corrections destroy the agreement with experimental 
data, especially, since the $\eta'$ mass is treated as a small quantity which
is, in our opinion, phenomenologically not justified. As mentioned before,
we treat the $\eta'$ as a massive state and thus the situation is similar to 
the case with baryons in conventional ChPT. In order to establish a chiral 
counting scheme even in the presence of massive states, a new regularization
scheme has been proposed in \cite{Becher:1999he} following the ideas of \cite{Tang:1996ca,Ellis:1998kc}.
The so-called infrared regularization keeps Lorentz and chiral invariance
explicit at all stages of the calculation and is thus perfectly suited
for $U(3)$ ChPT. The underlying idea of this regularization scheme is to 
extract the infrared singularities of the loop graphs in a relativistically 
invariant fashion. The chiral expansion of the remainder of the integral is an 
ordinary Taylor series in the momenta and Goldstone boson masses and is 
absorbed into the parameters of the counterterms which are renormalized 
accordingly. This means that the finite pieces of the infrared part of the
integral are kept whereas the remainder is omitted.

\begin{figure}
\centering
\includegraphics{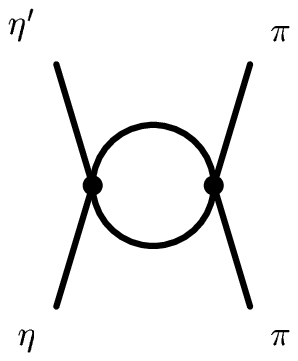}\hfill
\includegraphics{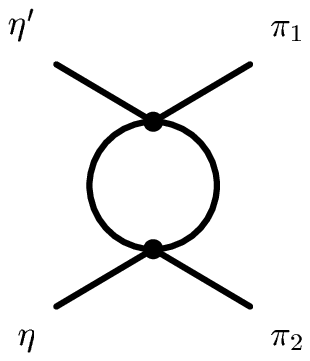}\hfill
\includegraphics{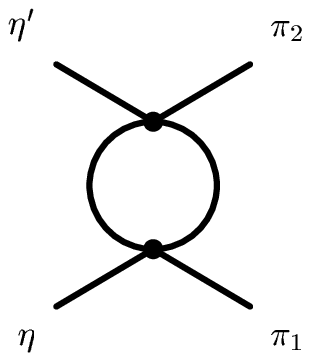}\hfill
\includegraphics{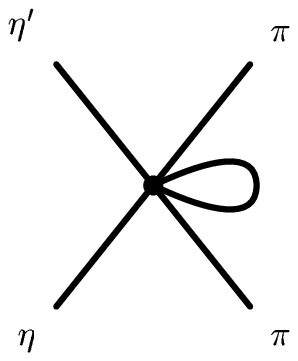}
\caption{$s$, $t$, $u$ and tadpole loop diagrams contributing to the decay.
         Solid lines denote the pseudoscalar mesons.}
\label{fig:loops}
\end{figure}

Infrared regularization has already been applied within $U(3)$ ChPT in
\cite{Borasoy:2001ik, Beisert:2001qb} where the resulting chiral series appeared to converge. In both
investigations only tadpoles contributed at one-loop order. In order to make a
more profound statement about the convergence of the chiral expansion , it is
necessary to consider unitarity corrections. Indeed, for the decay process 
considered in this paper both tadpoles and one-loop diagrams of the self-energy
type contribute, see Fig.~\ref{fig:loops}, and we will calculate the leading chiral loops, 
i.e. the leading chiral corrections from loops with the vertices
$\cvtwid{2}{1}, \beta_{5,18}$ and $\beta_{0,3,13}$.

The results for the fundamental integrals have already been presented in 
\cite{Becher:1999he} which we briefly discuss here restricting ourselves to the 
calculation of the finite pieces and neglecting the divergences.
The finite infrared parts of the tadpoles are readily evaluated
\begin{eqnarray}\label{eq:chilog}
\tad{a}  &=& 
\int_I \frac{d^d l}{(2 \pi)^d} \frac{i}{l^2 - m_a^2 +i\epsilon}
 = \frac{m_a^2}{16 \pi^2} \,  \ln \frac{m_a^2}{\mu^2} \, ;
\qquad a = \pi, K, \eta \nonumber \\
\tad{\eta'}  &=& 
\int_I \frac{d^d l}{(2 \pi)^d} \frac{i}{l^2 - m_{\eta'}^2 +i \epsilon}= 0
\end{eqnarray}
where the subscript $I$ denotes the infrared portion of the integral and $\mu$
is the scale introduced in infrared regularization.
The vanishing of the $\eta'$ tadpole is due to the lack of a low-energy scale
within the integral, such as the Goldstone boson masses.

The fundamental scalar integral with two propagators reads
\begin{equation}  \label{eq:integ}
\bonbon{p^2}{ab} = 
\int_I \frac{d^d l}{(2 \pi)^d} 
\frac{\I}{\big((p-l)^2 - m_a^2 +i \epsilon\big)\big(l^2 - m_b^2 +i \epsilon\big)}.
\end{equation}
If both mesons inside the loop are Goldstone bosons the infrared part is
obtained by regularizing the integral dimensionally. 
Expressed in terms of elementary functions the regularized 
integral equals
\begin{eqnarray}
\bonbon{p^2}{ab}=&&
\frac{1}{16\pi^2}\bigg[-1+
\ln\frac{m_{a} m_{b}}{\mu^2}
  +\frac{m^2_a-m^2_b}{p^2}\ln\frac{m_a}{m_b}
\nonumber\\&&\qquad\qquad\quad\mathord{}
  -\frac{2\sqrt{\Delta}}{p^2}\artanh\frac{\sqrt{\Delta}}{(m_a+m_b)^2-p^2}\bigg]
\nonumber\\
\Delta=&&\big((m_a-m_b)^2-p^2\big)\big((m_a+m_b)^2-p^2\big).
\end{eqnarray}
Outside their real-valued domains ($x>0$) the involved functions may be replaced 
according to
\begin{equation}
\sqrt{-x}=\I\sqrt{x},\quad
\artanh \I x=\I\arctan x,\quad
\artanh (-1-x)=\sfrac{1}{2}\I\PI+\arcoth (-1-x).
\end{equation}
For a loop with an $\eta'$ and a Goldstone boson the pertinent result reads
$(a= \pi, K, \eta)$
\begin{eqnarray}   \label{eq:eprimeloop}
\bonbon{p^2}{\eta'a}
  =&&    \frac{1}{16 \pi^2}\, 
\frac{\alpha (\Omega +\alpha)}{1 + 2 \alpha \Omega + \alpha^2} 
( \ln \alpha^2 -1) \nonumber 
\\&&\mathord{}
+\frac{1}{16 \pi^2}\,
\frac{\alpha \sqrt{\Omega^2 -1}}{1 + 2 \alpha \Omega + 
\alpha^2}  \ln \frac{\Omega +\alpha + \sqrt{\Omega^2 -1}}{\Omega +\alpha - 
\sqrt{\Omega^2 -1}}
\end{eqnarray}
where $\alpha = m_a/m_{\eta'}$ and $\Omega = (p^2 - m_{\eta'}^2 - m_a^2)/(2
m_{\eta'} m_a)$ is counted as a zeroth order quantity, $\Omega =
\mathcal{O}(q^0)$, and we have omitted the divergent pieces. 
The result \eqref{eq:eprimeloop} is valid for $\Omega^2 \ge 1$;
for $\Omega^2 < 1$ analytic continuation yields
\begin{eqnarray}   
\bonbon{p^2}{\eta'a}  
=&&\frac{1}{16 \pi^2} 
\frac{\alpha (\Omega +\alpha)}{1 + 2 \alpha \Omega + \alpha^2} 
(\ln \alpha^2 -1)\nonumber 
\\&&\mathord{}
+\frac{1}{8 \pi^2}\frac{\alpha \sqrt{1-\Omega^2}}{1+2\alpha\Omega+\alpha^2}  
\arccos \Big(-\frac{\Omega+\alpha}{\sqrt{1+2\alpha\Omega+\alpha^2}}\Big)
\end{eqnarray}
which is the result presented in \cite{Becher:1999he}.

The chiral expansion of this expression has been shown to converge in the
interval \cite{Becher:1999he}
\begin{equation}
- \frac{1 + \alpha^2}{2 \alpha} < \Omega < \frac{1 + \alpha^2}{2 \alpha} . 
\end{equation}
Since we do not consider isospin violation, a loop with an $\eta$ and $\eta'$
can only occur in the $s$ channel. In this case, $\Omega$ is constrained by
\begin{equation}
-1 \ge \Omega > -1.09  > - \frac{1 + \alpha^2}{2 \alpha} \approx -1.16 .
\end{equation}
In the $t$ and $u$ channels the $\eta'$ propagates only in combination with a 
pion and again the constraint for the limits is fulfilled
\begin{equation}
-1 \ge \Omega > -1.77 > - \frac{1 + \alpha^2}{2 \alpha} \approx -3.54 .
\end{equation}
Hence, in both possible cases with exactly one $\eta'$ inside the loop the 
chiral series converges and we can expand the expression \eqref{eq:eprimeloop} in 
$\alpha$ at fixed $\Omega$, obtaining $\bonbon{p^2}{\eta'a} = 
\mathcal{O}(\alpha)$. 
In comparison with the pure Goldstone boson loop
which is of chiral order $\mathcal{O}(1)$, the case
with one $\eta'$ propagator starts contributing at one chiral order higher.
In other words, the leading chiral contributions from the loops are
the ones with two Goldstone bosons, while those with an $\eta'$ and
a Goldstone boson are suppressed by one chiral order.
It is thus beyond our working precision of taking only the leading chiral
loops into account and we can safely neglect these contributions.

Finally, in the $s$ channel two $\eta'$ mesons can propagate inside the loop.
This integral was shown in \cite{Becher:1999he} to contain no infrared singularities and
we will omit these loops as well.
In order to calculate the leading loop corrections, it is sufficient
to consider the pure Goldstone boson loops, while dropping those with the
propagation of an $\eta'$.
To the order we are working the $\eta'$ field can therefore be regarded as an 
external field which does not appear in loops. This argument can be turned 
around by treating the $\eta'$ field as an external field right from the 
beginning and considering Goldstone boson loops only.
In the present calculation effects of the $\eta'$ within the loops appear at
sub-leading order whereas they are omitted completely in the latter approach.

In App. \ref{sec:ampli} we have presented the amplitudes
including the leading non-analytic contributions 
in infrared regularization.
The terms $A_2$ and $A_4$ are the tree diagrams from 
the corresponding Lagrangians $\Lagr^{(2)}$ and $\Lagr^{(4)}$, $A_Z$ stems from the 
wave function renormalization of the octet fields,
$A_{mf}$ is due to the higher order corrections of the meson masses and
the pion decay constant
and $A_\Delta, A_{s,t,u}$ are the tadpole and the unitarity corrections
in the $s,t,u$-channel, respectively.
We have presented the amplitudes for the bare unmixed fields
$\eta_0$ and $\eta_8$, e.g. $A_{88\pi\pi}=A_{88\pi\pi}^{(2)}+A_{88\pi\pi}^{(4)}$ is the expression for two external
$\eta_8$ fields. In order to obtain the amplitude for the physical fields
$\eta$ and $\eta'$, one must include $\eta$-$\eta'$ mixing and wave-function renormalization
of the $\eta_0$. 
It would then be sufficient 
to calculate $A_{08\pi\pi}$ up to fourth chiral order and both $A_{88\pi\pi}$ 
and $A_{00\pi\pi}$ only at lowest order 
$A_{88\pi\pi}^{(2)}$, $A_{00\pi\pi}^{(2)}$
and take $\eta$-$\eta'$
mixing into account. However, this result turns out to be scale dependent both 
at second and fourth chiral order due to $s, t, u \sim \mathcal{O}(1)$. 
The scale dependence at second order
is removed by calculating the complete one-loop amplitude 
$A^{(4)}_{88\pi\pi}$
as given in App. \ref{sec:ampli} and applying $\eta$-$\eta'$ mixing at second order.
The residual scale dependence is then of fourth chiral order and proportional
to the mixing angle. It could in principle
be cancelled by including, among other contributions, $\eta$-$\eta'$ mixing at fourth order
which is a two-loop effect and beyond the accuracy of this calculation.
We will therefore content ourselves by removing the scale dependence at second chiral order
and leave the scale dependence at fourth order which is numerically less significant.
Note that the amplitude $A_{00\pi\pi}$ does not exhibit any scale dependence up
to fourth chiral order and it suffices to include only the lowest order contribution 
$A_{00\pi\pi,2}^{(2)}$.
The complete amplitude we are using is given by
\begin{equation}\label{eq:Anlo}
A\indup{NLO}=\big(1+R^{(2)}_{0\eta'}\big)A_{08\pi\pi}+
R^{(2)}_{8\eta'}\big(A_{88\pi\pi}-R^{(2)}_{8\eta}A_{88\pi\pi,2}\big)
+R^{(2)}_{0\eta}A_{00\pi\pi,2} .
\end{equation}
The factor $R^{(2)}_{0\eta'}$ accounts for the $Z$-factor of the $\eta'$ field,
$R^{(2)}_{8\eta'}, R^{(2)}_{0\eta}$ are due to  $\eta$-$\eta'$ mixing,
and $R^{(2)}_{8\eta}$ corrects for the $Z$-factor of the $\eta_8$, 
since it was applied to both  $\eta_8$ fields in $A_{88\pi\pi}$.
Using the rough estimates for the parameters $\beta_i$ as explained above
we obtain at the central point in phase space, $(s_0,t_0,u_0)$,
for $\mu = m_{\eta'}$ 
$A\indup{NLO}(\mu = m_{\eta'})=-5.1+0.22\I$
and for $\mu = m_{\rho}$ the result reads 
$A\indup{NLO}(\mu = m_{\rho})=-4.4+0.22i$.
The values of the non-zero LECs $\cbeta{i}{0}$ were shifted according to 
their renormalization constants to accomodate for a modified scale $\mu$.

However, a fit to the experimental value of the decay width can
easily be accommodated by varying the LECs within their 
phenomenological ranges, e.g. 
$\beta_{0,3,13}$ can be modified while keeping
the other couplings fixed. 
The values of ($\beta_{0,3,13}=1.4$ at $\mu=m_{\rho}$)
or
($\beta_{0,3,13}=1.3$ at $\mu=m_{\eta'}$)
lead both to $A\indup{NLO}=-6.6+0.22\I$.
Such a fit is not unique 
since it could also be accomodated by varying some of the other LECs within their 
penomenological ranges and/or finetuning the OZI suppressed corrections
and we will refrain from performing one.
Furthermore, some of 
the higher order corrections (see below) 
tend to increase the magnitude of the amplitude cancelling
partially the effect of $A\indup{NLO}$.

Clearly, the loop corrections using the couplings from the leading
Lagrangian $\Lagr^{(2)}$ are under control with a scale dependence which is
considerably smaller than the uncertainty of the parameters $\beta_i$, see
\eqref{eq:estimate}.


\subsection{Loops from the fourth order lagrangian}

In the preceding section, we have included loop contributions with vertices
from the lowest order Lagrangian $\Lagr^{(2)}$. However, certain terms
from the  Lagrangian $\Lagr^{(4)}$ appear to spoil the strict chiral counting
scheme due to $s, t, u \sim \mathcal{O}(1)$ and contribute at lower chiral
orders as one would naively expect.
In fact, under reasonable assumptions for the LECs the $\beta_{0,3,13}$ 
term dominates the other tree level contributions by an order of magnitude.
It is therefore a legitimate question to ask, whether loop diagrams with
vertices from $\Lagr^{(4)}$, in particular $\beta_{0,3,13}$, are numerically
under control and do not upset the convergence of the amplitude expanded in terms of
derivatives and Goldstone boson masses. In this section we will clarify the issue
by calculating one-loop diagrams with 
$\cbeta{0}{0}$, $\cbeta{4}{0}$ and $\cbeta{5}{0}$ vertices,
which were part of the $\beta_{5,18}, \beta_{0,3,13}, 
\cbtwid{4}{1}$ and $\cbtwid{5}{1}$ terms we have included in the leading
order amplitude $A\indup{LO}$, \eqref{eq:a02}.
Strictly speaking, they are of fourth chiral order in the derivative/meson mass 
expansion of the effective Lagrangian. This implies that the one-loop diagrams 
with these vertices have eight powers in the external momenta and/or meson 
masses. At this order many more terms contribute, especially counterterms which 
compensate their scale dependence. A complete calculation up to eighth chiral
order, however, is far beyond the scope of this investigation and we will
content ourselves by presenting numerical results at the scales
$\mu=m_{\rho}=770\MeV$ and $\mu=m_{\eta'}=958\MeV$.
The dependence of the non-analytic pieces on the scale might also give a hint
on the importance of neglected LECs. 
Numerical results are given in Tab. \ref{tab:HighLoop}.

\begin{table}
\[
\begin{array}{|c||cc|cc|}\hline
\mathrm{diag.}&A_\rho&\tilde\alpha_\rho&A_{\eta'}&\tilde\alpha_{\eta'}\\\hline\hline
                         v&\,\,+1.4+0.0\I\,\,&\,\,+0.00+0.00\I\,\,&\,\,+1.4+0.0\I\,\,&\,\,+0.00+0.00\I\,\,\\
                 \beta_{0}&-8.6+0.0\I&-0.19+0.00\I&-8.6+0.0\I&-0.19+0.00\I\\
                 \beta_{4}&+1.4+0.0\I&+0.12+0.00\I&+1.4+0.0\I&+0.12+0.00\I\\
                 \beta_{5}&-1.1+0.0\I&-0.00+0.00\I&-1.1+0.0\I&+0.00+0.00\I\\
              v\mathrm{-}v&+0.9+0.2\I&-0.03+0.03\I&+1.7+0.2\I&-0.07+0.03\I\\\hline
        \beta_0\mathrm{-}v&-0.1-1.0\I&-0.09-0.14\I&+0.0-1.0\I&-0.11-0.14\I\\
        \beta_4\mathrm{-}v&-0.1+0.2\I&-0.09-0.20\I&-0.6+0.2\I&+0.25-0.20\I\\
        \beta_5\mathrm{-}v&-0.2-0.2\I&+0.02-0.02\I&+0.6-0.2\I&+0.06-0.02\I\\
\beta_0\mathrm{-}\beta_{0}&-0.0-0.2\I&+0.01+0.00\I&+0.1-0.2\I&+0.02-0.00\I\\
\beta_0\mathrm{-}\beta_{4}&-0.0+0.1\I&-0.51-0.30\I&+0.1+0.1\I&-0.76-0.30\I\\
\beta_0\mathrm{-}\beta_{5}&+0.2-0.0\I&-0.05-0.01\I&-0.2-0.0\I&-0.08-0.01\I\\
\beta_4\mathrm{-}\beta_{4}&-0.1-0.0\I&-0.08+0.00\I&-0.0-0.0\I&-0.19+0.09\I\\
\beta_4\mathrm{-}\beta_{5}&+1.1+0.0\I&+0.18-0.06\I&+1.8+0.0\I&+0.60-0.05\I\\
\beta_5\mathrm{-}\beta_{5}&-1.4-0.0\I&+0.03-0.00\I&-2.9-0.0\I&+0.06+0.00\I\\\hline
\end{array}
\]
\caption{
Amplitudes from higher loops. ``$a$-$b$'' denotes a loop between an $a$ and a $b$ vertex. 
``$v$'' denotes any vertex from the second order Lagrangian. For comparison 
we give the tree level amplitudes ``$a$'' of the considered vertices $a$.
$A$ gives the amplitude at the central point 
in phase space $s_0,t_0,u_0$ \eqref{eq:center}
and $\tilde\alpha=(-1/\sqrt{42})(\partial A/\partial y)$ 
corresponds roughly to the slope parameter $\alpha$ of \eqref{eq:me}.
The loop integrals are evaluated at the scales $\mu=m_\rho$ and $\mu=m_{\eta'}$.
}
\label{tab:HighLoop}
\end{table}

Although the tree level contributions from the $\beta_{0,3,13}$ term dominate
the amplitude, the loop corrections 
turn out to be much smaller indicating the convergence of the chiral expansion.
The same is true for the tadpole loops of fourth order couplings, they were
found to be an order of magnitude smaller than the corresponding tree
diagrams.
A sample analytical expressions for the one-loop diagrams is relegated to 
App. \ref{sec:loopamp}.

\section{Comparison with experiment}\label{sec:compare}
Our numerical results for the decay widths must be compared with the 
experimental ones
\begin{eqnarray} 
\Gamma(\eta' \rightarrow \eta \pi^+ \pi^-)  &=& (90 \pm 8)\keV 
\nonumber \\
\Gamma(\eta' \rightarrow \eta \pi^0 \pi^0)  &=& (42 \pm 4)\keV
\end{eqnarray}
The pion and kaon masses are taken to be
$\mmass{\pi}=138\MeV$, $\mmass{K}=496\MeV$
and the pion decay constant is $\decaypi=92.4\MeV$.

The large uncertainty in the LECs, in particular for $\beta_{0,3,13}$, prevents us
from making sensible numerical predictions. We have rather used the experimental
decay width to pin down the LEC combination $\beta_{0,3,13}$ more accurately.
Taking the expression $A\indup{LO}$ up to second order \eqref{eq:a02} one obtains
$\beta_{0,3,13}=(1.1\pm 0.4)\times 10^{-3}$ to be compared with the value
$\beta_{0,3,13}=(1.2\pm 1.1)\times 10^{-3}$ from conventional ChPT.
The error bars could be reduced considerably while the central values are
consistent. At fourth order in the chiral expansion of the decay amplitude
loops enter together with new coupling constants. A fit to the decay width is 
possible but not unique.
Further information is contained in the energy dependence of the decay
which will provide a better test for our approach. 
To this end, two Dalitz variables $x$ and $y$ are introduced with
\begin{eqnarray} 
x &=& \frac{\sqrt{3}}{Q} ( E_{\pi1} - E_{\pi2} )
\nonumber \\
y &=&  - \frac{2 + m_\eta/m_\pi}{Q} ( E_{\pi1} + E_{\pi2} ) -1
+ \frac{2 + m_\eta/m_\pi}{Q} ( m_{\eta'} - m_\eta)
\end{eqnarray}
with $Q = m_{\eta'} - m_\eta - 2 m_\pi$ and $E_{\pi i}$ being the energies
of the pions. As $E_{\pi 1}$ and $E_{\pi 2}$ vary over the 
physical region, $x$ ranges from about $-1.4$ to $1.4$ and $y$ ranges from $-1$
to about $1.2$.
Experimentally, the decay of the $\eta'$ is then found to be described by 
the matrix element
\begin{equation}  \label{eq:me}
|A|^2 = b \big( 1+ 2\alpha y+a y^2 + c x^2 \big)
\end{equation}
and fits to the data lead to
$\alpha=- 0.08 \pm 0.03 $ for $\eta' \rightarrow \eta \pi^+ \pi^-$
\cite{Kalbfleisch:1974ku} and $\alpha=- 0.058 \pm 0.013$, $a=\alpha^2 \pm 0.13^2$, 
$c=0.00 \pm 0.03$  for $\eta' \rightarrow \eta \pi^0 \pi^0$
\cite{Alde:1986nw}. More recently the decay  $\eta' \rightarrow \eta \pi^+ \pi^-$ was
used as a normalization for rare $\eta'$ decay searches and found to deliver
$\alpha = - 0.021 \pm 0.025$ \cite{Briere:1999bp}, a value consistent with, but
smaller than previous experiments. \footnote{The linear fit in Fig.~4 of 
\cite{Briere:1999bp}, however, shows a slope of $\alpha=-0.038$, in
contradiction with the value quoted and closer
to the value of \cite{Alde:1986nw}.}

These experimental data must be compared with our calculations. 
We will keep the notation of \eqref{eq:me}. 
Since the $\beta_{0,3,13}$ contribution dominates the amplitude,
it is instructive to investigate its impact on the parameters in 
\eqref{eq:me}. We do this by keeping only the $\beta_{0,3,13}$ term in the
amplitude while neglecting all other contributions. With these
simplifications $b$ in \eqref{eq:me} is fixed by the value of
$\beta_{0,3,13}$ and the dependence of $A$ on the Mandelstam variables
determines the remaining parameters. 
The values of these slope parameters are functions of 
the masses of the four involved particles only
and we obtain
$\alpha = -0.138$, $a=-0.0016$, $c=-0.082$
which turn out to be 
larger in magnitude than the experimental data. 
Albeit we are able to explain the measured decay width with the
$\beta_{0,3,13}$ term only, it does not reproduce the value for $\alpha$. 

Let us therefore focus on the effect
of further terms in $A\indup{LO}$  on the slope $\alpha$. 
Most of those terms in App. \ref{sec:b} are constant in phase space and
influence $\alpha$ indirectly since $\alpha$ is the \emph{relative} slope.
The only operators besides $\beta_{0,3,13}$ yielding a sizeable slope are
the $\cbtwid{4}{1}$ and $\cbtwid{5}{1}$ terms 
(the contribution from the $88\pi\pi$ amplitude is suppressed due to mixing.)
We find that the relative slope 
($\alpha/A$ in Tab. \ref{tab:HighLoop}) of these terms is
opposite and much larger than that of $\beta_{0,3,13}$. 
Therefore these terms reduce $\alpha$ without
affecting the decay width. 
By setting $\cbtwid{4}{1}=\cbeta{4}{0}=-0.7\times 10^{-3}$,
$\cbtwid{5}{1}=\cbeta{5}{0}=1.9\times 10^{-3}$
and $\beta_{0,3,13}=1.1\times 10^{-3}$ and all others at their central values
our findings for the slope parameters resulting from $A\indup{LO}$ are
\begin{equation}\label{eq:coeffme1}
\Gamma=42.3\keV,\qquad\alpha=-0.057,\qquad a=-0.022,\qquad c=-0.10.
\end{equation}
Including the chiral corrections from $A\indup{NLO}$ \eqref{eq:Anlo} some of the
LECs must be modified within their phenomenological ranges to restore
the experimental decay width. Such a fit to the decay width is not unique.
One possibility is to increase the value of $\beta_{0,3,13}$ in order to
bring $A\indup{NLO}$ to agreement with the experimental decay width, 
while keeping the other couplings fixed. The slope parameters read then 
\begin{equation}\label{eq:coeffme2}
\Gamma=43.6\keV, \qquad\alpha=-0.113,\qquad a=-0.019,\qquad c=-0.19
\end{equation}
at the scale $\mu =m_\eta'$
but one should keep in mind that $\alpha$ depends strongly on 
the choice of parameters and no precise statement about its value can be
made at this level.
E.g. by modifying some of the operators constant in phase
space in order to accommodate the decay width instead of increasing 
$\beta_{0,3,13}$ -- the resulting $|\alpha|$ tends to be smaller than the
value shown in \eqref{eq:coeffme2}.

\begin{figure}
\centering

\includegraphics{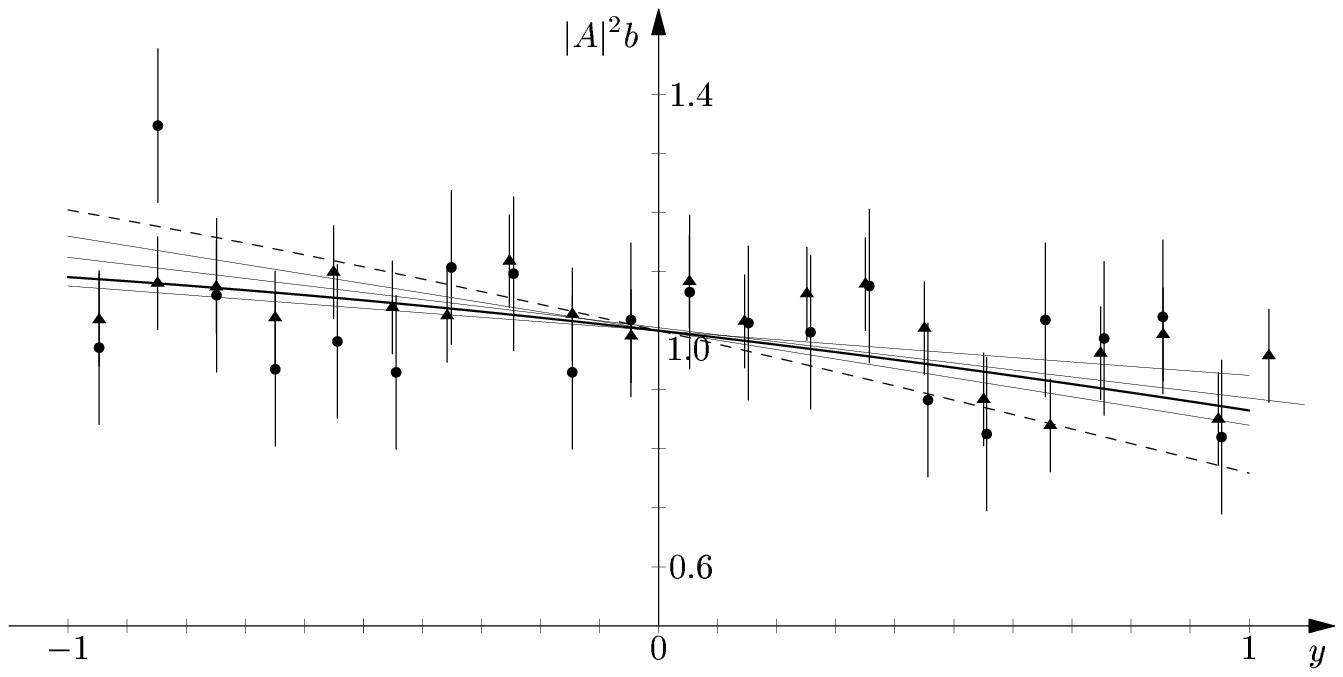}
\bigskip\bigskip

\includegraphics{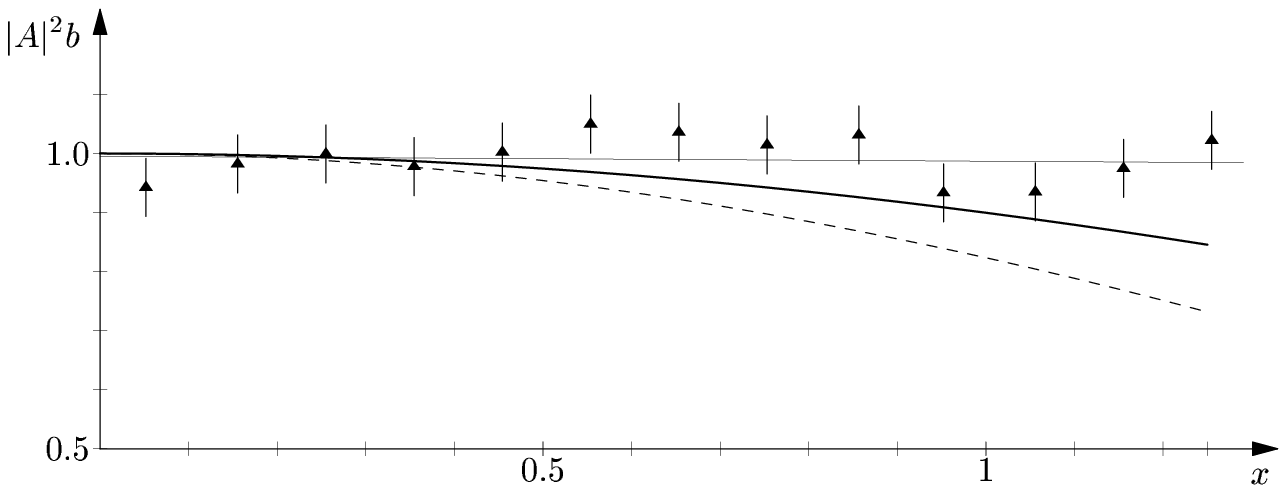}

\caption{Upper (lower) graph: 
The dependence of $|A|^2$ on the Dalitz variable $y$ ($x$).
The thick solid line denotes our $A\indup{LO}$ result, and
the dashed line represents the theoretical uncertainty due to the chiral
corrections from $A\indup{NLO}$.
Triangles and circles are the experimental 
data from \cite{Alde:1986nw} and \cite{Briere:1999bp}, respectively.
The thin lines are, descending in magnitude of slope, 
the linear fits from 
\cite{Kalbfleisch:1974ku}, \cite{Alde:1986nw} and \cite{Briere:1999bp} (\cite{Alde:1986nw}).}
\label{fig:dalitzy}
\end{figure}

Under reasonable assumptions on the appearing LECs, such as neglecting OZI 
violating corrections to the couplings,
we are thus able to achieve reasonable agreement with experimental data.
Hence,  \eqref{eq:coeffme1} constitutes our central values
for the slope parameters and \eqref{eq:coeffme2} is a 
conservative estimate of their uncertainties from $A\indup{NLO}$
related to the choice of the LECs. We observe a tendency to larger
values of $|\alpha|$ if $A\indup{NLO}$ is taken into account. A non-vanishing
slope $|\alpha|$ may represent the contribution of a gluon component to the
$\eta'$ decay \cite{Alde:1986nw}.
Fig.~\ref{fig:dalitzy} shows the dependence 
of $|A|^2$ on both, $x$ and $y$,
together with the experimental data from \cite{Kalbfleisch:1974ku}, \cite{Alde:1986nw} and \cite{Briere:1999bp}.

Furthermore,
the leading order contributions to the slope $\alpha$ are due to
counterterms of the fourth order chiral Lagrangian. In order to obtain a
more reliable value for $\alpha$, one must evaluate next-to-leading order
chiral corrections including terms from the sixth order Lagrangian which
we do not consider here.

Nevertheless, it would be preferable to have an independent confirmation
of the value for $\beta_{0}$. Such a check is provided by the calculation
of scalar meson resonances using a coupled channel approach \cite{Beisert:2001A1}.
In the present context, it is sufficient to sketch briefly some of the
results, for details the reader is referred to \cite{Beisert:2001A1}. In $SU(3)$
ChPT the resonances $f_0(980)$ and $a_0(980)$ have been reproduced using
coupled channels and taking only the second order Lagrangian into account
\cite{Oller:1997ti}. In \cite{Beisert:2001A1} we applied a similar coupled channel
analysis but using $U(3)$ ChPT and with the inclusion of the fourth order Lagrangian.
While the established resonances at $980\MeV$ remained unchanged, 
it gave rise to additional resonances in the isoscalar and isovector
channels which could be identified with 
the $f_0(1370)$ or $f_0(1500)$ and 
$a_0(1450)$, respectively.
This result is similar to \cite{Oller:1998hw} where the fourth 
order Lagrangian had to be included to obtain the 
vector resonance $\rho$.
Our results for the partial decay widths of the isovector
agree with the experimental data of $a_0$. The counterterms of the
fourth order Lagrangian with four derivatives, such as the $\beta_{0}$ operator,
turned out to be essential in producing the new resonances.
In particular we found that the resonances were 
not seen for $\beta_{0,3,13}\lesssim 0.5\times 10^{-3}$ and did not
fit the experiment for $\beta_{0,3,13}\gtrsim 1.5\times 10^{-3}$.
The other fourth order operators only changed the parameters slightly.
This is actually not surprising since for momenta well beyond the scale of
chiral symmetry breaking around $1\GeV$ momentum dependent
terms will dominate (assuming they are comparable to other terms
at smaller scales).
In particular, the range for $\beta_{0,3,13}$ is consistent with the value given 
in \eqref{eq:b0est}
and this confirms the choice we made for $\beta_{0,3,13}$.

Finally, we would like to comment on final state interactions due to 
two-pion rescattering as they have been evaluated, e.g., for the
decay $\eta\to\pi\pi\pi$ \cite{Kambor:1996yc}. The final state interaction
of two pions in the $I=0$, $s$-wave
channel is strong and attractive and is believed to dominate
interactions between the $\eta$ and a pion in 
$\eta'\to\eta\pi\pi$.
In \cite{Kambor:1996yc} the Khuri-Treiman equations were used to determine numerically 
the unitarity corrections. The corrections to the decay amplitude 
were found to be $14\%$
at the center of the decay region.
We note that the energy of the two pion state is approximately the
same
for the decay $\eta'\to\eta\pi\pi$ and roughly estimate the two-pion
final state interactions for the present decay to be of similar 
size, i.e. less than $20\%$. Comparing this with, e.g., the uncertainty 
of the dominant coupling $\beta_{0,3,13}$ \eqref{eq:b0est} the corrections
are moderate and smaller than the given uncertainty for $\beta_{0,3,13}$.
Hence final state interactions do not modify our conclusions considerably and
we can safely neglect them.

\section{Summary and Conclusions}\label{sec:conclude}

In the present work, we have calculated the decay 
$\eta' \rightarrow \eta \pi\pi$ within the framework of infrared regularized 
$U(3)$ ChPT. We presented
the most general effective Lagrangian up to fourth order in the derivative 
expansion which describes the interactions at low energies between the members
of the lowest lying pseudoscalar nonet, i.e. the octet of Goldstone bosons
($\pi, K,\eta$) and the corresponding singlet state $\eta'$. At first sight,
the fourth order Lagrangian looks less encouraging due to the proliferation of
new unknown LECs. However, the unknown parameters entering the calculation 
can be identified either as OZI violating corrections
of known LECs or couplings of parity-odd operators. 
Assuming that their contribution is negligible we are left with 
already known LECs and borrow their values from conventional $SU(3)$ ChPT.

The tree diagram contributions including $\eta$-$\eta'$ mixing have been
evaluated first and it was found that together with the LEC combination
$\cvtwid{2}{1}$ from the second order Lagrangian, the two combinations
$\beta_{5,18}$ and $\beta_{0,3,13}$ from the fourth order Lagrangian
constitute the decay amplitude up to second order in the Goldstone boson 
masses. In fact, it turns out that $\beta_{0,3,13}$ dominates the decay 
amplitude and exceeds by an order of magnitude the contribution from 
$\cvtwid{2}{1}$. 
The decay width 
 can be used to reduce the uncertainty for
$\beta_{0,3,13}$ yielding
$\beta_{0,3,13}=(1.1\pm 0.4)\times 10^{-3}$ 
to be compared with $\beta_{0,3,13}=(1.2\pm 1.1)\times 10^{-3}$ from conventional ChPT.
The dominance of $\beta_{0,3,13}$ has been explained in the
light of Adler zeros. Similar results at the tree level were obtained
within large $N_c$ ChPT \cite{Herrera-Siklody:1999ss} but loops were omitted since they are a 
next-to-next-to-leading order effect in the combined 1/$N_c$ and chiral 
expansion. In contradistinction to large $N_c$ ChPT we treat the $\eta'$ as a 
heavy field and evaluate loop diagrams by employing infrared regularization 
which keeps Lorentz and chiral invariance explicit at all stages of the 
calculation but suppresses contributions from loops with an $\eta'$. Within this scenario
loop contributions start already at next-to-leading order.
We restrict ourselves to the calculation of one-loop diagrams -- both tadpoles
and unitarity corrections -- with vertices from $\Lagr^{(2)}$ and the LECs
$\cbeta{0}{0}$, $\cbeta{4}{0}$ and $\cbeta{5}{0}$. In infrared regularization
loop diagrams with a virtual $\eta'$ contribute only at higher orders and
are therefore neglected. We find that the Goldstone boson loop contributions, 
in particular those with a $\beta_{0,3,13}$ vertex, are rather small indicating
the convergence of the chiral series. Reasonable agreement with experimental
data is obtained and both the decay width and its energy dependence on the
Dalitz variables are well reproduced. However, our results are subject to the choice of parameters
such as the omission of OZI violating couplings.

Overall we can state, that infrared regularized $U(3)$ ChPT is suited
to study the $\eta'$ decay $\eta' \rightarrow \eta \pi \pi$. We were able to reproduce the 
experimental data of the decay 
by taking into account next-to-leading order effects including one-loop 
corrections. Our results indicate furthermore the convergence of the chiral
series.
In the present work we did not discuss the possibility of including low lying
scalar mesons, especially the $a_0(980)$, as has been done in a tree level 
model, see \cite{Fariborz:1999gr}. In our approach the explicit inclusion of these
resonances is not necessary in order to obtain agreement with experiment
although their effects might be hidden in some LECs. The $a_0(980)$ resonance,
e.g., does indeed contribute to some of the couplings $\beta_i$ in the 
resonance exchange picture \cite{Ecker:1989te}.

\section*{Acknowledgements}

We would like to thank Stefan Wetzel for reading the manuscript and Norbert 
Kaiser for valuable discussions.

\appendix 

\section{Abbreviations} \label{sec:b}

This is a summary of LEC combinations as they appear in the text:
\begin{eqnarray}
\beta_{0,3,13}&=&\cbeta{0}{0}+\cbeta{3}{0}-\sfrac{3}{4}\cbeta{13}{0}
\nonumber\\
\beta_{4,5,17,18}&=&3\cbeta{4}{0}+\cbeta{5}{0}-9\cbeta{17}{0}+3\cbeta{18}{0}
\nonumber\\
\beta_{5,18}&=&\cbeta{5}{0}+\sfrac{3}{2}\cbeta{18}{0}
\nonumber\\
\cvtwid{2}{1}&=&\sfrac{1}{4}\decaypi^2-\sfrac{1}{2}\sqrt{6}\coeffv{3}{1}
\nonumber\\
\cvtwid{2}{2}&=&\sfrac{1}{4}\decaypi^2-\sqrt{6}\coeffv{3}{1}-3\coeffv{2}{2}
\nonumber\\
\cbtwid{4}{1}&=&\cbeta{4}{0}+\sfrac{1}{2}\sqrt{6}\cbeta{22}{1}
\nonumber\\
\cbtwid{5}{1}&=&\cbeta{5}{0}+\sfrac{1}{2}\sqrt{6}\cbeta{21}{1}
\nonumber\\
\cbtwid{26}{1}&=&\cbeta{6}{0}+\cbeta{7}{0}-\sfrac{1}{4}\sqrt{6}\cbeta{26}{1}
\nonumber\\
\cbtwid{8}{1}&=&\cbeta{8}{0}-\sfrac{1}{4}\sqrt{6}\cbeta{25}{1}
\end{eqnarray}

\section{Amplitudes} \label{sec:ampli}

In this appendix, we present the amplitudes before applying 
$\eta$-$\eta'$ mixing and $\eta'$ wave-function renormalization. As explained
in the main text, the amplitude can then be divided into three pieces
$A_{08\pi\pi}, A_{88\pi\pi}$ and $A_{00\pi\pi}$ with the subscripts denoting
the external fields. 
The terms $A_2$ and $A_4$ are the tree diagrams from 
the corresponding Lagrangians $\Lagr^{(2)}$ and $\Lagr^{(4)}$.
The terms $A_Z,A_{mf}$ stem from the 
wave-function renormalization of the octet fields and
the expansion of meson masses and decay constants
and $A_\Delta, A_{s,t,u}$ are the tadpole and the unitarity corrections
in the $s,t,u$-channel, respectively.
The fourth order amplitude
$A^{(4)}_{88\pi\pi}$ has been included in order
to reduce the scale dependence, although it contributes via $\eta$-$\eta'$ mixing 
to the $\eta'$ decay at sixth order in the derivative expansion.

\begin{eqnarray}
A^{(2)}_{08\pi\pi}&=&
\frac{4\sqrt{2}\,\mmass{\pi}\cvtwid{2}{1}}{3\decaypi^4}
\nonumber\\
A^{(4)}_{08\pi\pi} &=& A_4+A_Z+A_{mf}+A_\Delta+A_s+A_t+A_u
\nonumber\\
A_{4}&=&\mathord{}
\frac{4\sqrt{2}\,\beta_{0,3,13}}{3\decaypi^4}
\big(s^2+t^2+u^2-2\mmass[4]{\pi}-\mmass[4]{\eta}-\mmass[4]{\eta'}\big)
-\frac{8\sqrt{2}\,\mmass{\pi}\mmass{\eta'} \beta_{5,18}}{3\decaypi^4}
\nonumber\\&&\mathord{}
+\frac{16\sqrt{2}\,(\mmass{K}-\mmass{\pi})\cbtwid{4}{1}}{3\decaypi^4}
\big(2\mmass{\pi}-s\big)
+\frac{4\sqrt{2}\,\mmass{\pi}\cbtwid{5}{1}}{3\decaypi^4}
\big(2\mmass{\eta'}-s-t-u\big)
\nonumber\\&&\mathord{}
+\frac{8\sqrt{2}\,\mmass{\pi}\cbtwid{26}{1}}{\decaypi^4}
\big(3\mmass{\pi}-\mmass{\eta}\big)
+\frac{64\sqrt{2}\,\mmass[4]{\pi}\cbtwid{8}{1}}{3\decaypi^4}
\nonumber\\
A_{Z}&=&\mathord{}
\frac{4\sqrt{2}\,\mmass{\pi}\cvtwid{2}{1}}{3\decaypi^6}
\Big[
-12 (2 m_K^2 + m_\pi^2) \cbeta{4}{0}
-4\big(\mmass{\eta}+2\mmass{\pi}\big)\cbeta{5}{0}
+\sfrac{2}{3}\tad{\pi}
+\sfrac{5}{6}\tad{K}
\Big]
\nonumber\\
A_{mf}&=&\mathord{}
\frac{4\sqrt{2}\,\mmass{\pi}\cvtwid{2}{1}}{3\decaypi^6}
\Big[
-16\big(\cbeta{6}{0}-\cbeta{4}{0}\big)(2 m_K^2 + m_\pi^2)
-16\big(\cbeta{8}{0}-\cbeta{5}{0}\big)\mmass{\pi}
-\sfrac{5}{2}\tad{\pi}-\tad{K}+\sfrac{1}{6}\tad{\eta}\Big]
\nonumber\\&&\mathord{}
-
\frac{4\sqrt{3}\,\mmass{\pi}\coeffv{3}{1} }{3\decaypi^6}
\big(8\cbeta{5}{0}\mmass{\pi}
+8\cbeta{4}{0}(2 m_K^2 + m_\pi^2)
-2\tad{\pi}-\tad{K}\big)
\nonumber\\
A_{\Delta}&=&\mathord{}
\frac{2\sqrt{2}\,\cvtwid{2}{1}\tad{K}}{45\decaypi^6}
\big(2\mmass{K}-7\mmass{\pi}\big)
-\frac{2\sqrt{2}\,\mmass{\pi}\cvtwid{2}{1}}{9\decaypi^6}
\big(\tad{\eta}+5\tad{\pi}\big)
\nonumber\\
A_{s}&=&\mathord{}
\frac{8\sqrt{2}\,\mmass{\pi}\cvtwid{2}{1}\tad{\pi}}{9\decaypi^6}
+\frac{2\sqrt{2}\,\cvtwid{2}{1}\tad{K}}{9\decaypi^6}
\big(\mmass{\pi}-2\mmass{K}\big)
\nonumber\\&&\mathord{}
+\frac{4\sqrt{2}\,\mmass{\pi}\cvtwid{2}{1}\bonbon{s}{\pi\pi}}{9\decaypi^6}
\big(2\mmass{\pi}-3s-\mmass{\pi}\big)
\nonumber\\&&\mathord{}
+\frac{\sqrt{2}\,\cvtwid{2}{1}s\,\bonbon{s}{KK}}{3\decaypi^6}
\big(2\mmass{K}-\mmass{\pi}\big)
\nonumber\\&&\mathord{}
+\frac{2\sqrt{2}\,\mmass{\pi}\cvtwid{2}{1}\bonbon{s}{\eta\eta}}{9\decaypi^6}
\big(2\mmass{\eta}-\mmass{\pi}\big)
\nonumber\\
A_{t}&=&\mathord{}
\frac{2\sqrt{2}\,(2 m_K^2 + m_\pi^2)\cvtwid{2}{1}\tad{K}}{9\decaypi^6}
-\frac{4\sqrt{2}\,\mmass[4]{\pi}\cvtwid{2}{1}\bonbon{t}{\eta\pi}}{9\decaypi^6}
\nonumber\\&&\mathord{}
+\frac{\sqrt{2}\,(2 m_K^2 + m_\pi^2)\cvtwid{2}{1}\bonbon{t}{KK}}{9\decaypi^6}
\big(3\mmass{\eta}+\mmass{\pi}-3t\big)
\nonumber\\
A_{u}&=&\mathord{}
\frac{2\sqrt{2}\,(2 m_K^2 + m_\pi^2)\cvtwid{2}{1}\tad{K}}{9\decaypi^6}
-\frac{4\sqrt{2}\,\mmass[4]{\pi}\cvtwid{2}{1}\bonbon{u}{\eta\pi}}{9\decaypi^6}
\nonumber\\&&\mathord{}
+\frac{\sqrt{2}\,(2 m_K^2 + m_\pi^2)\cvtwid{2}{1}\bonbon{u}{KK}}{9\decaypi^6}
\big(3\mmass{\eta}+\mmass{\pi}-3u\big).
\end{eqnarray}

\begin{eqnarray}
A^{(2)}_{88\pi\pi}&=&
\frac{\mmass{\pi}}{3\decaypi^2}
\nonumber\\
A^{(4)}_{88\pi\pi} &=& A_4+A_Z+A_{mf}+A_\Delta+A_s+A_t+A_u
\nonumber\\
A_{4}&=&\mathord{}
\frac{4\big(\cbeta{0}{0}+\cbeta{3}{0}\big)}{3\decaypi^4}
\big(s^2+t^2+u^2-2\mmass[4]{\pi}-\mmass[4]{\eta}-\mmass[4]{\eta'}\big)
+\frac{4\mmass{\pi}\cbeta{5}{0}}{3\decaypi^4}\big(s+t+u\big)
\nonumber\\&&\mathord{}
+\frac{8\cbeta{1}{0}}{\decaypi^4}
\big(s-\mmass{\eta}-\mmass{\eta'}\big)\big(s-2\mmass{\pi}\big)
-\frac{128\mmass{\pi}(\mmass{K}-\mmass{\pi})\cbeta{7}{0}}{3\decaypi^4}
+\frac{64\mmass[4]{\pi}\cbeta{8}{0}}{\decaypi^4}
\nonumber\\&&\mathord{}
+\frac{4\cbeta{2}{0}}{\decaypi^4}
\Big[\big(t-\mmass{\pi}-\mmass{\eta'}\big)\big(t-\mmass{\pi}-\mmass{\eta}\big)
+\big(u-\mmass{\pi}-\mmass{\eta'}\big)\big(u-\mmass{\pi}-\mmass{\eta}\big)\Big]
\nonumber\\&&\mathord{}
+\frac{8\cbeta{4}{0}}{\decaypi^4}
\Big[\mmass{\eta}\big(s-2\mmass{\pi}\big)
+\mmass{\pi}\big(s-\mmass{\eta}-\mmass{\eta'}\big)\Big]
+\frac{8\mmass{\pi}\cbeta{6}{0}}{\decaypi^4}
\big(5\mmass{\eta}+\mmass{\pi}\big)
\nonumber\\
A_{Z}&=&\mathord{}
\frac{\mmass{\pi}}{3\decaypi^4}
\Big[
-16\cbeta{4}{0}(2 m_K^2 + m_\pi^2)
-8\cbeta{5}{0}\big(\mmass{\eta}+\mmass{\pi}\big)
+\sfrac{2}{3}\tad{\pi}
+\sfrac{4}{3}\tad{K}
\Big]
\nonumber\\
A_{mf}&=&
\frac{\mmass{\pi}}{3\decaypi^4}
\Big[-16\big(\cbeta{8}{0}-\cbeta{5}{0}\big)\mmass{\pi}
-16\big(\cbeta{6}{0}-\cbeta{4}{0}\big)(2 m_K^2 + m_\pi^2)
-\sfrac{5}{2}\tad{\pi}-\tad{K}+\sfrac{1}{6}\tad{\eta}\Big]
\nonumber\\
A_{\Delta}&=&\mathord{}
\frac{\tad{K}}{90\decaypi^4}\big(32\mmass{K}-4\mmass{\pi}-3s-3t-3u\big)
-\frac{\mmass{\pi}}{18\decaypi^4}\big(\tad{\eta}+5\tad{\pi}\big)
\nonumber\\
A_{s}&=&\mathord{}
\frac{2\mmass{\pi}\tad{\pi}}{9\decaypi^4}
+\frac{\tad{K}}{36\decaypi^4}\big(4\mmass{\pi}+18s-3t-3u\big)
\nonumber\\&&\mathord{}
+\frac{\mmass{\pi}\bonbon{s}{\pi\pi}}{6\decaypi^4}
\big(\mmass{\pi}-2s\big)
+\frac{\mmass{\pi}\bonbon{s}{\eta\eta}}{18\decaypi^4}
\big(\mmass{\pi}-4\mmass{\eta}\big)
\nonumber\\&&\mathord{}
+\frac{s\,\bonbon{s}{KK}}{24\decaypi^4}
\big(2\mmass{\pi}+3\mmass{\eta}+3\mmass{\eta'}-9s\big)
\nonumber\\
A_{t}&=&\mathord{}
\frac{\tad{K}}{36\decaypi^4}\big(3s+18t+3u-12\mmass{\eta}\big)
-\frac{\mmass[4]{\pi}\bonbon{t}{\eta\pi}}{9\decaypi^4}
\nonumber\\&&\mathord{}
-\frac{\bonbon{t}{KK}}{36\decaypi^4}
\big(2\mmass{\eta}+\mmass{\pi}+\mmass{\eta'}-3t\big)
\big(3\mmass{\eta}+\mmass{\pi}-3t\big)
\nonumber\\
A_{u}&=&\mathord{}
\frac{\tad{K}}{36\decaypi^4}\big(3s+3t+18u-12\mmass{\eta}\big)
-\frac{\mmass[4]{\pi}\bonbon{u}{\eta\pi}}{9\decaypi^4}
\nonumber\\&&\mathord{}
-\frac{\bonbon{u}{KK}}{36\decaypi^4}
\big(2\mmass{\eta}+\mmass{\pi}+\mmass{\eta'}-3u\big)
\big(3\mmass{\eta}+\mmass{\pi}-3u\big).
\end{eqnarray}

\bigskip\bigskip\bigskip

\begin{equation}\label{eq:A00pp}
A^{(2)}_{00\pi\pi}=\frac{8\cvtwid{2}{2}\mmass{\pi}}{3\decaypi^4}
+\frac{4\coeffv{1}{2}}{\decaypi^4}\big(\mmass{\pi}+\mmass{\pi}-s\big)
\end{equation}

\newpage

\section{A sample result for a higher order loop} \label{sec:loopamp}
In this appendix, we present a sample result for the higher order loops which 
involve vertices from the Lagrangian $\Lagr^{(4)}$. We have chosen the
contribution from a loop with two $\cbeta{5}{0}$ vertices. 
The other contributions are obtained in a similar way but turn out to be
rather lengthy. We have therefore omitted them for brevity.
\begin{eqnarray}
A_s&=&\mathord{}
\frac{8\sqrt{2}\big(\cbeta{5}{0}\big)^2}{81 \decaypi^8}\Big[
-6\tad{\pi}\mmass[4]{\pi}\big(52\mmass{K}+221\mmass{\pi}+39\mmass{\eta'}-33s\big)
\nonumber\\&&\quad\mathord{}
+6\tad{K}\big(
52\mmass[6]{K}
+191\mmass[4]{K}\mmass{\pi}
-141\mmass{K}\mmass[4]{\pi}
-18\mmass[6]{\pi}
\nonumber\\&&\quad\quad\quad\mathord{}
+3\mmass{\eta'}(3\mmass{K}-2\mmass{K})(\mmass{\pi}+3\mmass[4]{\pi})
+3s(9\mmass[4]{\pi}-11\mmass[4]{K})\big)
\nonumber\\&&\quad\mathord{}
+2\tad{\eta}\mmass{\pi}\big(8\mmass{K}-5\mmass{\pi}\big)
\big(20\mmass{K}+\mmass{\pi}+3\mmass{\eta}+3s\big)
\nonumber\\&&\quad\mathord{}
-12\bonbon{s}{\pi\pi}\mmass[4]{\pi}
\big(4\mmass{K}+5\mmass{\pi}+3\mmass{\eta'}\big)
\big(13\mmass{\pi}-6s\big)
\nonumber\\&&\quad\mathord{}
+3\bonbon{s}{KK}\big(
2\mmass[4]{K}+12\mmass{K}\mmass{\pi}+2\mmass[4]{\pi}
-3s(\mmass{K}+\mmass{\pi})\big)
\nonumber\\&&\quad\quad\quad\mathord{}
\times\big(
28\mmass[4]{K}-19\mmass{K}\mmass{\pi}
+3\mmass{\eta'}(3\mmass{K}-2\mmass{\pi})
-9s(\mmass{K}-\mmass{\pi})
\big)
\nonumber\\&&\quad\mathord{}
+4\bonbon{s}{\eta\eta}\mmass{\pi}
\big(8\mmass{K}-5\mmass{\pi}\big)
\big(2\mmass{K}+\mmass{\pi}\big)
\big(4\mmass{K}-\mmass{\pi}+\mmass{\eta'}\big)
\Big]
\nonumber\\
A_t&=&\mathord{}
\frac{8\sqrt{2}\big(\cbeta{5}{0}\big)^2}{81\decaypi^8}\Big[
-6\tad{\pi}\mmass[4]{\pi}\big(8\mmass{K}+13\mmass{\pi}+3\mmass{\eta'}-3t\big)
\nonumber\\&&\quad\mathord{}
-6\tad{K}\big(
73\mmass[4]{K}\mmass{\pi}
+43\mmass{K}\mmass[4]{\pi}
+10\mmass[6]{\pi}
\nonumber\\&&\quad\quad\quad\mathord{}
+3\mmass{\eta'}
(\mmass{K}+2\mmass{\pi})
(\mmass{K}+\mmass{\pi})
+9t(\mmass[4]{K}-4\mmass{K}\mmass{\pi}-\mmass[4]{\pi})\big)
\nonumber\\&&\quad\mathord{}
-6\tad{\eta}\mmass[4]{\pi}\big(16\mmass{K}+5\mmass{\pi}
+3\mmass{\eta'}+3t\big)
\nonumber\\&&\quad\mathord{}
-8\bonbon{t}{\eta\pi}\mmass[4]{\pi}
\big(4\mmass{K}+5\mmass{\pi}+3\mmass{\eta'}\big)
\big(2\mmass{K}+\mmass{\pi}\big)
\nonumber\\&&\quad\mathord{}
-\bonbon{t}{KK}
\big(
46\mmass[4]{K}+16\mmass{K}\mmass{\pi}+10\mmass[4]{\pi}
-9t(5\mmass{K}+\mmass{\pi})\big)
\nonumber\\&&\quad\quad\quad\mathord{}
\times\big(9\mmass{K}\mmass{\pi}
+\mmass{\eta'}(2\mmass{\pi}+\mmass{K})
+3t(\mmass{K}-\mmass{\pi})\big)
\Big]
\end{eqnarray}
%


\bibliography{etapipi} 

\begin{thebibliography}{10}
\raggedright

\bibitem{DiVecchia:1980ve}
P.~Di~Vecchia and G.~Veneziano,
\textit{``Chiral dynamics in the large n limit''},
Nucl. Phys. \textbf{B171} (1980) 253.

\bibitem{Witten:1980sp}
E.~Witten,
\textit{``Large n chiral dynamics''},
Ann. Phys. \textbf{128} (1980) 363.

\bibitem{Gasser:1985gg}
J.~Gasser and H.~Leutwyler,
\textit{``Chiral perturbation theory: Expansions in the mass of the strange
  quark''},
Nucl. Phys. \textbf{B250} (1985) 465.

\bibitem{Leutwyler:1996sa}
H.~Leutwyler,
\textit{``Bounds on the light quark masses''},
Phys. Lett. \textbf{B374} (1996) 163,
\href{http://arXiv.org/abs/hep-ph/9601234}{\texttt{hep-ph/9601234}}.

\bibitem{Herrera-Siklody:1997pm}
P.~Herrera-Sikl\`ody, J.~I. Latorre, P.~Pascual and J.~Taron,
\textit{``Chiral effective {Lagrangian} in the large-{$N_c$} limit: The nonet
  case''},
Nucl. Phys. \textbf{B497} (1997) 345,
\href{http://arXiv.org/abs/hep-ph/9610549}{\texttt{hep-ph/9610549}}.

\bibitem{Borasoy:2000xj}
B.~Borasoy,
\textit{``The {$\eta'$} and the topological charge density''},
Eur. Phys. J. \textbf{A7} (2000) 255,
\href{http://arXiv.org/abs/hep-ph/0002165}{\texttt{hep-ph/0002165}}.

\bibitem{Kaiser:1998ds}
R.~Kaiser and H.~Leutwyler,
\textit{``Pseudoscalar decay constants at large {$N(c)$}''},
\href{http://arXiv.org/abs/hep-ph/9806336}{\texttt{hep-ph/9806336}}.

\bibitem{Kaiser:2000gs}
R.~Kaiser and H.~Leutwyler,
\textit{``Large {$N(c)$} in chiral perturbation theory''},
Eur. Phys. J. \textbf{C17} (2000) 623,
\href{http://arXiv.org/abs/hep-ph/0007101}{\texttt{hep-ph/0007101}}.

\bibitem{Becher:1999he}
T.~Becher and H.~Leutwyler,
\textit{``Baryon chiral perturbation theory in manifestly lorentz invariant
  form''},
Eur. Phys. J. \textbf{C9} (1999) 643,
\href{http://arXiv.org/abs/hep-ph/9901384}{\texttt{hep-ph/9901384}}.

\bibitem{Borasoy:2001ik}
B.~Borasoy and S.~Wetzel,
\textit{``{$U(3)$} chiral perturbation theory with infrared regularization''},
Phys. Rev. \textbf{D63} (2001) 074019,
\href{http://arXiv.org/abs/hep-ph/0105132}{\texttt{hep-ph/0105132}}.

\bibitem{Beisert:2001qb}
N.~Beisert and B.~Borasoy,
\textit{``{$\eta$}-{$\eta'$} mixing in {$U(3)$} chiral perturbation theory''},
Eur. Phys. J. \textbf{A11} (2001) 329,
\href{http://arXiv.org/abs/hep-ph/0107175}{\texttt{hep-ph/0107175}}.

\bibitem{Belkov:1987ms}
A.~A. Bel'kov and V.~N. Pervushin,
\textit{``Chiral {$p(4)$} {Lagrangians} and amplitude of {$\eta'\to\eta 2 \pi$}
  decay''},
Sov. J. Nucl. Phys. \textbf{45} (1987) 891.

\bibitem{Fajfer:1989ij}
S.~Fajfer and J.~M. G\'erard,
\textit{``Hadronic decays of {$\eta$} and {$\eta'$} in the large n limit''},
Z. Phys. \textbf{C42} (1989) 431.

\bibitem{Akhoury:1989as}
R.~Akhoury and M.~Leurer,
\textit{``Low-energy effective {Lagrangian} description of {$\eta$} and
  {$\eta'$} decays''},
Z. Phys. \textbf{C43} (1989) 145.

\bibitem{Herrera-Siklody:1999ss}
P.~Herrera-Sikl\`ody,
\textit{``{$\eta$} and {$\eta'$} hadronic decays in {$U(3)_L \times U(3)_R$}
  chiral perturbation theory''},
\href{http://arXiv.org/abs/hep-ph/9902446}{\texttt{hep-ph/9902446}}.

\bibitem{Fariborz:1999gr}
A.~H. Fariborz and J.~Schechter,
\textit{``{$\eta'\to\eta\pi\pi$} decay as a probe of a possible lowest-lying
  scalar nonet''},
Phys. Rev. \textbf{D60} (1999) 034002,
\href{http://arXiv.org/abs/hep-ph/9902238}{\texttt{hep-ph/9902238}}.

\bibitem{Schechter:1971tc}
J.~Schechter and Y.~Ueda,
\textit{``General treatment of the breaking of chiral symmetry and scale
  invariance in the {$SU(3)$} sigma model''},
Phys. Rev. \textbf{D3} (1971) 2874.

\bibitem{Singh:1975aq}
C.~A. Singh and J.~Pasupathy,
\textit{``On the decay modes of the meson {$\eta'(958)$} and chiral symmetry
  breaking''},
Phys. Rev. Lett. \textbf{35} (1975) 1193,
Erratum-ibid. \textbf{35} (1975) 1748.

\bibitem{Deshpande:1978iv}
N.~G. Deshpande and T.~N. Truong,
\textit{``Resolution of the {$\eta'\to\eta\pi\pi$} puzzle''},
Phys. Rev. Lett. \textbf{41} (1978) 1579.

\bibitem{Bramon:1980ni}
A.~Bramon and E.~Masso,
\textit{``On the quark content of {$\Delta(980)$} and other scalar mesons''},
Phys. Lett. \textbf{B93} (1980) 65,
Erratum-ibid. \textbf{B107} (1980) 455.

\bibitem{Castoldi:1988dm}
P.~Castoldi and J.~M. Fr\`ere,
\textit{``{$\eta'\to\pi^+\pi^-\pi^0$}: A key to understanding the
  {$\eta$}-{$\eta'$} system''},
Z. Phys. \textbf{C40} (1988) 283.

\bibitem{Groom:2000in}
Particle Data Group Collaboration, D.~E. Groom et~al.,
\textit{``Review of particle physics''},
Eur. Phys. J. \textbf{C15} (2000) 1.

\bibitem{Bijnens:1994qh}
J.~Bijnens, G.~Ecker and J.~Gasser,
\textit{``Chiral perturbation theory''},
\href{http://arXiv.org/abs/hep-ph/9411232}{\texttt{hep-ph/9411232}}.

\bibitem{Balog:1984ps}
J.~Balog,
\textit{``Effective {Lagrangian} from {QCD} anomalies''},
Phys. Lett. \textbf{B149} (1984) 197.

\bibitem{Andrianov:1985ay}
A.~A. Andrianov,
\textit{``Bosonization in four-dimensions due to anomalies and an effective
  {Lagrangian} for pseudoscalar mesons''},
Phys. Lett. \textbf{B157} (1985) 425.

\bibitem{Espriu:1990ff}
D.~Espriu, E.~de~Rafael and J.~Taron,
\textit{``The {QCD} effective action at long distances''},
Nucl. Phys. \textbf{B345} (1990) 22,
Erratum-ibid. \textbf{B355} (1990) 278.

\bibitem{Bijnens:1993uz}
J.~Bijnens, C.~Bruno and E.~de~Rafael,
\textit{``{Nambu}-{Jona}-{Lasinio} like models and the low-energy effective
  action of {QCD}''},
Nucl. Phys. \textbf{B390} (1993) 501,
\href{http://arXiv.org/abs/hep-ph/9206236}{\texttt{hep-ph/9206236}}.

\bibitem{Ecker:1989te}
G.~Ecker, J.~Gasser, A.~Pich and E.~de~Rafael,
\textit{``The role of resonances in chiral perturbation theory''},
Nucl. Phys. \textbf{B321} (1989) 311.

\bibitem{Tang:1996ca}
H.-B. Tang,
\textit{``A new approach to chiral perturbation theory for matter fields''},
\href{http://arXiv.org/abs/hep-ph/9607436}{\texttt{hep-ph/9607436}}.

\bibitem{Ellis:1998kc}
P.~J. Ellis and H.-B. Tang,
\textit{``Pion nucleon scattering in a new approach to chiral perturbation
  theory''},
Phys. Rev. \textbf{C57} (1998) 3356,
\href{http://arXiv.org/abs/hep-ph/9709354}{\texttt{hep-ph/9709354}}.

\bibitem{Kalbfleisch:1974ku}
G.~R. Kalbfleisch,
\textit{``Comments on the {$\eta'(958)$}: Branching ratio, linear matrix
  element and dipion phase shift''},
Phys. Rev. \textbf{D10} (1974) 916.

\bibitem{Alde:1986nw}
Serpukhov-Brussels-Los Alamos-Annecy (LAPP) Collaboration, D.~Alde et~al.,
\textit{``Matrix element of the {$\eta'(958)\to\eta\pi^0\pi^0$} decay''},
Phys. Lett. \textbf{B177} (1986) 115.

\bibitem{Briere:1999bp}
CLEO Collaboration, R.~A. Briere et~al.,
\textit{``Rare decays of the {$\eta'$}''},
Phys. Rev. Lett. \textbf{84} (2000) 26,
\href{http://arXiv.org/abs/hep-ex/9907046}{\texttt{hep-ex/9907046}}.

\bibitem{Beisert:2001A1}
N.~Beisert and B.~Borasoy,
\textit{``Anomalous decays''},
work in progress.

\bibitem{Oller:1997ti}
J.~A. Oller and E.~Oset,
\textit{``Chiral symmetry amplitudes in the s-wave isoscalar and isovector
  channels and the {$\sigma$}, {$f_0(980)$}, {$a_0(980)$} scalar mesons''},
Nucl. Phys. \textbf{A620} (1997) 438,
Erratum-ibid. \textbf{A652} (1997) 407,
\href{http://arXiv.org/abs/hep-ph/9702314}{\texttt{hep-ph/9702314}}.

\bibitem{Oller:1998hw}
J.~A. Oller, E.~Oset and J.~R. Pelaez,
\textit{``Meson-meson and meson-baryon interactions in a chiral
  non-perturbative approach''},
Phys. Rev. \textbf{D59} (1999) 074001,
Erratum-ibid. \textbf{D60} (1999) 099906,
\href{http://arXiv.org/abs/hep-ph/9804209}{\texttt{hep-ph/9804209}}.

\bibitem{Kambor:1996yc}
J.~Kambor, C.~Wiesendanger and D.~Wyler,
\textit{``Final state interactions and {Khuri}-{Treiman} equations in {$\eta\to
  3\pi$} decays''},
Nucl. Phys. \textbf{B465} (1996) 215,
\href{http://arXiv.org/abs/hep-ph/9509374}{\texttt{hep-ph/9509374}}.

\end{thebibliography}
\bibliographystyle{nb}

\end{document}